\documentclass[%
 reprint, %linenumbers,
superscriptaddress,
 amsmath,amssymb,
 aps, physrev,
]{revtex4-2}

\pdfoutput=1

\usepackage{graphicx}% Include figure files
\usepackage{dcolumn}% Align table columns on decimal point
\usepackage{bm}% bold math
\usepackage{amsmath,mathrsfs,amssymb,dsfont}
\usepackage{xcolor}
\usepackage{import}
\usepackage{enumitem}
\usepackage[normalem]{ulem}
\usepackage{soul}
\usepackage{physics}
\usepackage{comment}
\usepackage{float}
\usepackage{afterpage}
\usepackage{pdfpages}
\usepackage{atbegshi}

\DeclareMathOperator{\orderof}{\sim}
\newcommand{\rr}{\mathbf{r}}

\newcommand{\ux}{\hat{\mathbf{x}}}
\newcommand{\uy}{\hat{\mathbf{y}}}
\newcommand{\uz}{\hat{\mathbf{z}}}
\newcommand{\id}{1\!\! 1}
\newcommand{\sinc}{\mathrm{sinc}}
\newcommand{\uvec}[1]{\hat{\mathbf{#1}}}

\makeatletter
\AtBeginDocument{\let\LS@rot\@undefined}
\makeatother

\begin{document}

\preprint{APS/123-QED}

\title{\textbf{Long-Lived Mechanically-Detected Molecular Spins for Quantum Sensing} 
}% 

\author{Sahand Tabatabaei}
\affiliation{Department of Physics and Astronomy, University of Waterloo, Waterloo, ON, Canada, N2L3G1}
\affiliation{Institute for Quantum Computing, University of Waterloo, Waterloo, ON, Canada, N2L3G1}

\author{Pritam Priyadarsi}
\affiliation{Department of Physics and Astronomy, University of Waterloo, Waterloo, ON, Canada, N2L3G1}
\affiliation{Institute for Quantum Computing, University of Waterloo, Waterloo, ON, Canada, N2L3G1}

\author{Daniel Tay}
\affiliation{Department of Physics and Astronomy, University of Waterloo, Waterloo, ON, Canada, N2L3G1}
\affiliation{Institute for Quantum Computing, University of Waterloo, Waterloo, ON, Canada, N2L3G1}

\author{Namanish Singh}
\affiliation{Department of Physics and Astronomy, University of Waterloo, Waterloo, ON, Canada, N2L3G1}
\affiliation{Institute for Quantum Computing, University of Waterloo, Waterloo, ON, Canada, N2L3G1}

\author{Pardis Sahafi}
\affiliation{Department of Physics and Astronomy, University of Waterloo, Waterloo, ON, Canada, N2L3G1}
\affiliation{Institute for Quantum Computing, University of Waterloo, Waterloo, ON, Canada, N2L3G1}

\author{Andrew Jordan}
\affiliation{Department of Physics and Astronomy, University of Waterloo, Waterloo, ON, Canada, N2L3G1}
\affiliation{Institute for Quantum Computing, University of Waterloo, Waterloo, ON, Canada, N2L3G1}

\author{Raffi Budakian}
\email{rbudakian@uwaterloo.ca}
\affiliation{Department of Physics and Astronomy, University of Waterloo, Waterloo, ON, Canada, N2L3G1}
\affiliation{Institute for Quantum Computing, University of Waterloo, Waterloo, ON, Canada, N2L3G1}

\date{\today}% It is always \today, today,

\begin{abstract}
Quantum sensors based on individual spins provide unprecedented access to local magnetic fields in condensed matter, chemistry, and biology, with solid-state defect spins emerging as the leading platform.
However, their molecular-sensing capabilities are limited by confinement to a host lattice, which prevents placement in close proximity to a target molecule.
Molecular spins offer an alternative, enabling chemical tunability and flexible positioning relative to the target system.
Here we present a nanoscale sensing platform that combines molecular electron spins, ultrasensitive mechanical readout, and Hamiltonian engineering.
Using a modified XYXY dipolar decoupling sequence, we suppress electron–electron dipolar interactions across a broad distribution of control fields, extending coherence times to $\orderof 400~\mu$s in an attoliter-scale droplet containing $\orderof$~100 trityl-OX063 radicals.
Leveraging this sequence, we demonstrate frequency-selective detection of nanotesla-scale AC fields and perform sensing and spectroscopy of small, local nuclear-spin ensembles.
Collectively, these results establish SQUINT (Spin-based QUantum Integrated Nanomechanical Transduction) as a framework for quantum sensing that affords molecular-level control over sensor properties and enables direct integration into complex molecular targets.
\end{abstract}
\maketitle

%\tableofcontents

%%%%%%%%%%%%%%%%%%%%% INTRODUCTION %%%%%%%%%%%%%%%%%%%%%%%%%%%%%%
\section{Introduction}

The ability to control and detect nuclear and electron spins via magnetic resonance has fundamentally transformed our understanding of molecular structure, condensed matter systems and chemistry, forming the bedrock of powerful techniques such as magnetic resonance imaging and spectroscopy.
Building on this foundation, developments in quantum technologies have re-imagined individual spins as nanoscale probes of their environment, capable of detecting magnetic fields \cite{Wolf2015, Hong2013, Barry2020}, temperature \cite{Neumann2013, Kucsko2013}, and strain \cite{Broadway2019, Ovartchaiyapong2014} with exceptional sensitivity.
Such quantum sensors have been proposed for extracting molecular-scale structural information from biomolecules and complex materials \cite{Kono2026,roadmap, Lovchinsky2016,Fazhan2015,Ajoy2015}, imaging quasiparticle excitations in condensed matter \cite{Casola2018,McCullian2020,vanderSar2015}, probing chemical processes \cite{Khurana2024, Finkler2021, Barton2020}, and even searching for physics beyond the standard model \cite{Chigusa2025,Jiao2021,Ebadi2022}.

Among existing platforms, nitrogen-vacancy (NV) centers in diamond \cite{Du2024, Schirhagl2014, roadmap} are a leading candidate, offering optical initialization and readout with long coherence times under ambient conditions.
These features have enabled landmark experiments ranging from magnetic resonance spectroscopy of single proteins \cite{Lovchinsky2016} to atomic-resolution imaging of nuclear spins within the diamond \cite{Zopes2018, Abobeih2019}.
Nonetheless, bringing the NV sensor close to the target is challenging, as shallow NV centers often have unstable charge states under optical illumination \cite{Bluvstein2019}, along with reduced coherence times due to magnetic and electric field noise near the diamond surface \cite{Ofori2012, Rosskopf2014,Kim2015}.
Despite ongoing efforts to mitigate these effects \cite{Bluvstein2019, Lo2025, Neethirajan2023, Zheng2022}, achieving robust NV operation within a few nanometers of the surface remains a persistent challenge, motivating alternative platforms such as defects in two-dimensional materials \cite{Roberts2025,Tetienne2021, Rizzato2023}. 
More generally, defect-based sensors are inherently constrained by their host crystals, limiting how sensor spins can be positioned or oriented relative to a target.
This has prompted interest in molecular spin systems as alternative quantum sensors whose placement and spin properties can be chemically engineered \cite{Yu2021, Bonizzoni2024, Singh2025}.

Molecular radicals are a particularly well-established class of spin systems that have long served as versatile probes in electron spin resonance (ESR) spectroscopy \cite{Jeschke2018, Jeschke2012, Duss2014}.
Decades of ESR research have yielded a comprehensive understanding of their spin dynamics, along with mature techniques for chemically synthesizing, functionalizing, and positioning radicals within a few nanometers of target molecules, either through site-directed attachment or by embedding them within the material under study \cite{Roser2016,Klare2013,Ketter2021,Tormyshev2020}.
These capabilities provide a degree of chemical tunability unique to molecular spin systems \cite{Yu2021}.
In addition, established cryoprotectants from structural biology enable low-temperature operation of the radicals while preserving the sample's structural integrity \cite{Malferrari2014}.
These features suggest that molecular radicals can serve as sensor spins placed in close and flexible arrangements relative to a target sample to interrogate its local magnetic environment.
However, their potential as quantum sensors has heretofore remained largely unexplored, in part because their coherence times are typically short---often less than $\orderof 10~\mu$s depending on the radical and its environment, even at low temperatures \cite{Jeschke2025,Savitsky2010,Kveder2008,Chen2016}.
Furthermore, although optically-addressable molecular spin systems are being developed \cite{Yu2021, Bayliss2020,Bayliss2022,Mena2024,Feder2025}, many existing ESR radicals---including nitroxides and trityls, the workhorses of modern structural biology---lack optical addressability.
This limitation motivates alternative approaches that can harness these established radicals for quantum sensing and potentially transform structural biology.

In this work, we introduce SQUINT (Spin-based QUantum Integrated Nanomechanical Transduction), a nanoscale quantum sensing platform that integrates molecular electron spins with ultrasensitive mechanical readout and Hamiltonian engineering.
SQUINT builds on experimental infrastructure developed for force-detected nanoscale magnetic resonance imaging (nanoMRI) \cite{roadmap}, enabling high-fidelity control and detection of nanoscale electron-spin ensembles while providing access to large, time-dependent magnetic-field gradients for high-resolution spatial encoding \cite{Tabatabaei2024, Haas2022, Rose2018}.

We use the electron spin in trityl-OX063 radicals as the quantum sensor.
To preserve coherence in these ensembles, we introduce a modified XYXY dipolar decoupling (XYXYd) sequence that suppresses electron–electron interactions, extending coherence times up to $\orderof 400~\mu$s.
Using this sequence, we demonstrate frequency-selective sensing of nanotesla-scale AC magnetic fields, and perform nuclear-spin sensing and spectroscopy of ${}^1$H and 1\% natural-abundance ${}^{13}$C spins on the OX063 molecule.
Sensitivity analysis indicates that near-term improvements to the mechanical readout could enable detection of individual ${}^1$H spins using a single electron spin at nanoscale separations.
Together, these results establish a route to leveraging widely used molecular radicals as practical quantum sensors for probing local magnetic environments in molecular systems.

%%%%%%%%%%%%%%%%%%%%% Sensing Platform %%%%%%%%%%%%%%%%%%%%%%%%%%%%%%
\section{Sensing Platform}

\subsection{The OX063 Molecular Quantum Sensor}

\begin{figure}[h!]
    \centering
    \includegraphics[width=\linewidth]{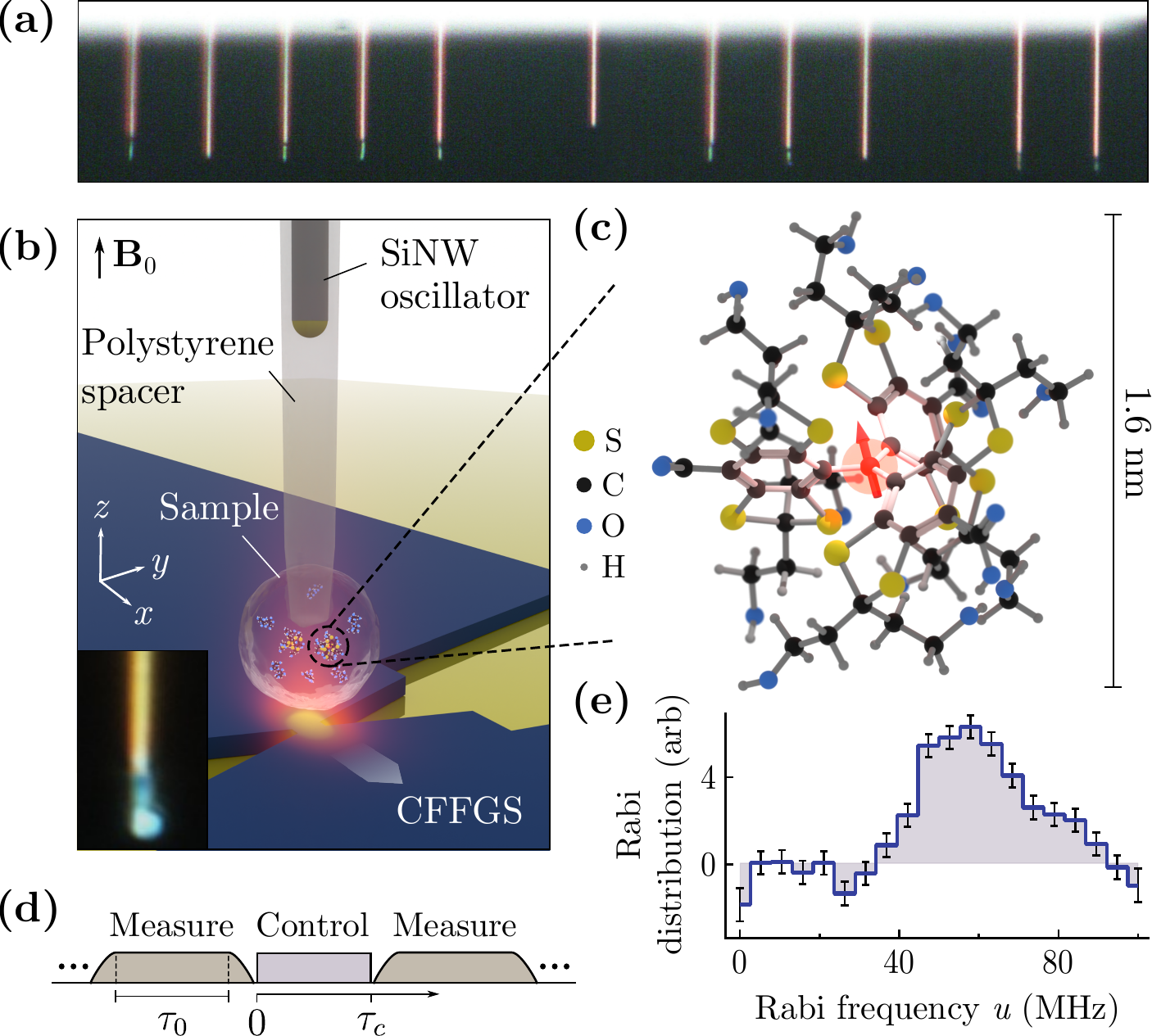}
    \caption{
        SQUINT sensing platform.
        (a) Optical image of the SiNW chip, showing an array of SiNW mechanical sensors fabricated along the chip edge.
        (b) Schematic of the sensing platform, including the SiNW oscillator, sample, polystyrene spacer, and current-focusing field gradient source (CFFGS).
        The bright red region indicates the spatial profile of the readout gradient.
        The molecules depicted inside the sample droplet are trehalose and trityl-OX063.
        Inset: Zoomed-in optical image of the tip of a SiNW, showing the polystyrene spacer and attached sugar droplet.
        (c) Molecular structure of the trityl-OX063 radical, with the central unpaired electron (red arrow) serving as the sensor spin.
        (d) Schematic of the correlation-based spin measurement protocol.
        The signal is obtained by correlating the readout-gradient-weighted longitudinal magnetization measured before and after a control sequence of duration $\tau_c$, with repeated measurement windows of duration $\tau_0$.
        Each measure block is comprised of the MAGGIC spin detection sequence \cite{Tabatabaei2024, Haas2022, Rose2018}.
        (e) Distribution of electron Rabi frequencies for the sample used in this work. The distribution was measured by driving the electron spins at the electron Larmor frequency with microwave pulses of variable encoding duration, following the Fourier encoding scheme described in \cite{Rose2018,Haas2022}. Each spin accumulates a rotation set by its local Rabi frequency, and a Fourier cosine transform of the measured signal gives the distribution of $u$ weighted by the readout gradient squared.
        The Rabi frequencies shown correspond to a peak current of $5$~mA through the CFFGS.
    }
    \label{fig1}
\end{figure}

Each trityl-OX063 molecule [Fig.~\ref{fig1}(c)] is a stable organic radical with a single unpaired electron primarily localized on the central carbon of the trityl core. In a static magnetic field, this spin ($S=1/2$) defines a coherently addressable two-level system and serves as the quantum sensor in SQUINT.
During a sensing sequence, variations in the local magnetic field modulate the electron-spin Zeeman splitting, producing phase accumulation of the spin coherence that is subsequently mapped onto longitudinal magnetization and read out mechanically.

In OX063, the unpaired electron is sterically shielded by three sulfur-substituted phenyl rings.
The shielding limits close contact between the radical center and surrounding spins, reducing unwanted magnetic interactions while preserving magnetic coupling to the local environment \cite{Lumata2013,Jeschke2025}.
Together with the high molecular symmetry of the radical, this protection gives OX063 a narrow ESR spectrum and small $g$-anisotropy compared with many other organic radicals \cite{Lumata2013}.
The peripheral carboxyl and hydroxyl groups improve water solubility and biocompatibility, and allow OX063 to be used in aqueous and glassy sugar cryoprotectant matrices \cite{Tormyshev2020,Brender2021}.

Functionalized OX063 derivatives also provide chemical handles for site-directed labeling of proteins and other target molecules \cite{Tormyshev2020,Ketter2021}. Taken together, these features make OX063 and related trityl radicals versatile ESR spin probes and, in this work, allow OX063 to serve as a stable, chemically adaptable electron-spin sensor embedded in the material under study.

\subsection{Sample and Spin Environment}

We prepare the sample as a 1–-10 aL droplet containing OX063 sensor spins dispersed in a 10:1 trehalose dihydrate:sucrose cryoprotectant mixture.
Such trehalose-based matrices are commonly used in structural biology, e.g. for ESR distance measurements, and can preserve molecular structure under low-temperature \cite{Malferrari2014,Meyer2015,Mathies2016} and low-hydration conditions \cite{Maiti2023,Luyckx2011}.

Previous measurements show that fluctuating dangling-bond spins on the surface of the silicon nanowire (SiNW) force sensor [Fig.~\ref{fig1}(b)] substantially shorten the OX063 electron-spin $T_1$ via excess relaxation \cite{Tabatabaei2024}.
To mitigate this effect, we mount the droplet at the end of a $\orderof 2-3~\mu$m-long polystyrene spacer attached to the SiNW tip, which separates the sensor spins from fast-relaxing paramagnetic defects on the nanowire.
With the spacer in place, we measure $T_1$ values of 500--700~ms, corresponding to about a tenfold increase over earlier results.
Details of the sample attachment are provided in the Supplemental Material \cite{Supplements}.

To improve experimental throughput, we prepared multiple SiNWs, each with an attached droplet sample, enabling sequential measurements of multiple samples within a single experimental run.
The results presented in this work correspond to one such sample, consisting of a $\sim 1$~aL detection volume containing an estimated 140 OX063 molecules.
Measurements were performed in high vacuum at a base temperature of 4~K under a static field $B_0 = 0.330$~T aligned with the SiNW axis, corresponding to an electron Larmor frequency of 9.26~GHz.
Owing to optical absorption from the interferometric displacement-detection laser, we estimate the SiNW temperature during measurement to be $\sim 8$~K.
Further details of the sample preparation are given in the Supplemental Material \cite{Supplements}.

\subsection{Spin Control and Mechanical Readout}

The SQUINT sensing platform, shown in Fig.~\ref{fig1}(b), follows the general configuration used in previous nanoMRI experiments \cite{Rose2018,Haas2022,Tabatabaei2021}.
A silicon nanowire (SiNW) nanomechanical oscillator detects the longitudinal spin magnetization by measuring the force on the electron spins in a magnetic-field gradient.
Spin control and detection are performed using a current-focusing field gradient source (CFFGS), which generates both the time-dependent gradients used for mechanical detection and the transverse fields used for spin control.
The CFFGS is broadband (DC--10~GHz) and can address both nuclear and electron spins. The silicon chip containing the SiNW array was positioned with piezo actuators so that the selected SiNW was aligned over the center of the CFFGS, with the attached sample approximately $70$~nm above the surface.

Driving the CFFGS at the spin Larmor frequency generates the transverse control (Rabi) field $B_1(\rr) = \sqrt{B_x^2(\rr) + B_y^2(\rr)}/2$, with $B_x(\rr),B_y(\rr),B_z(\rr)$ being the components of the magnetic field at location $\rr$.
Since the CFFGS field is spatially non-uniform, the resulting electron Rabi frequency $u(\rr) = \gamma_e B_1(\rr)/(2\pi)$ varies across the sample volume, where $\gamma_e/(2\pi) = 28.025$~GHz/T is the electron gyromagnetic ratio.
An example of this distribution, measured for the sample used in this work, is shown in Fig.~\ref{fig1}(e).

Readout of the sensor spins is conducted by detecting statistical correlations in their spin fluctuations using the modulated alternating gradients generated with currents (MAGGIC) protocol described in our previous work \cite{Tabatabaei2024, Haas2022, Rose2018}.
The detected signal is the correlation between the gradient-weighted longitudinal magnetization before and after a control sequence of interest [Fig.~\ref{fig1}(d)].
Rather than measuring an average magnetization as in conventional ESR, this correlation measurement detects statistical spin fluctuations in a maximally mixed ensemble.
The signal is proportional to $C\propto \sum_j G^2(\rr_j) \Tr[S_j^z(0) S_j^z(\tau_c)]$, where $G(\rr_j) = \partial B_z(\rr_j)/\partial y$ is the readout gradient at the position of the $j^\text{th}$ electron, $\tau_c$ is the duration of the control sequence, and $S_j^z(t)$ is the Heisenberg-picture $z$-spin operator.

%%%%%%%%%%%%%%%%%%%%% XYXYd Sequence %%%%%%%%%%%%%%%%%%%%%%%%%%%%%%
\section{XYXY\lowercase{d} Sequence}

Preserving the sensor-spin coherence requires a control sequence that averages out both local resonance offsets and electron-electron dipolar interactions, and operates over the broad Rabi frequency distribution generated by the CFFGS [Fig.~\ref{fig1}(e)].
For this purpose, we construct a modified XYXY dipolar decoupling (XYXYd) sequence based on specifically engineered adiabatic inversions that rescale the dipolar interaction.
In the high-field limit, the spin Hamiltonian in the electron rotating frame is $H(t) = H_\Delta + H_D + H_C(t)$, where
\begin{align}
    H_\Delta = -\sum_j \Delta_j S_j^z 
\end{align}
encompasses all electron spin resonance offsets, including hyperfine interactions to surrounding nuclei,
\begin{align}
    H_D
    = \sum_{j<k} D_{jk}\!\left(3 S_j^z S_k^z - \mathbf S_j\!\cdot\!\mathbf S_k\right)
\end{align}
is the secular electron-electron dipolar interaction, and
\begin{align}
    H_C(t) = 2\pi \sum_j u(\rr_j)~\mathbf{a}(t)\cdot\mathbf{S}_j
\end{align}
is the microwave control Hamiltonian.
The transverse vector $\mathbf a(t)$ satisfies 
$\vert \mathbf a(t)\vert\leq 1$ and determines the
instantaneous amplitude and phase of the applied field.

Conventional dynamical decoupling sequences based on trains of global $\pi$ pulses, such as XYXY \cite{Alvarez2010, AliAhmed2013, Wang2012}, refocus static frequency offsets described by $H_\Delta$ but leave the dipolar Hamiltonian $H_D$ unchanged to leading order.
For the electron-spin ensemble used here, we measured an inhomogeneous dephasing time of $T_2^* = 78(6)$~ns and a Hahn-echo coherence time of $T_2 = 6(1)~\mu$s, indicating that suppressing $H_\Delta$ alone is insufficient.
Achieving the longer coherence times reported in this work therefore requires a sequence that goes beyond conventional spin-echo decoupling by also averaging electron--electron dipolar interactions. This requirement motivates the XYXYd sequence developed here.

\begin{figure}[h!]
    \centering
    \includegraphics[width=\linewidth]{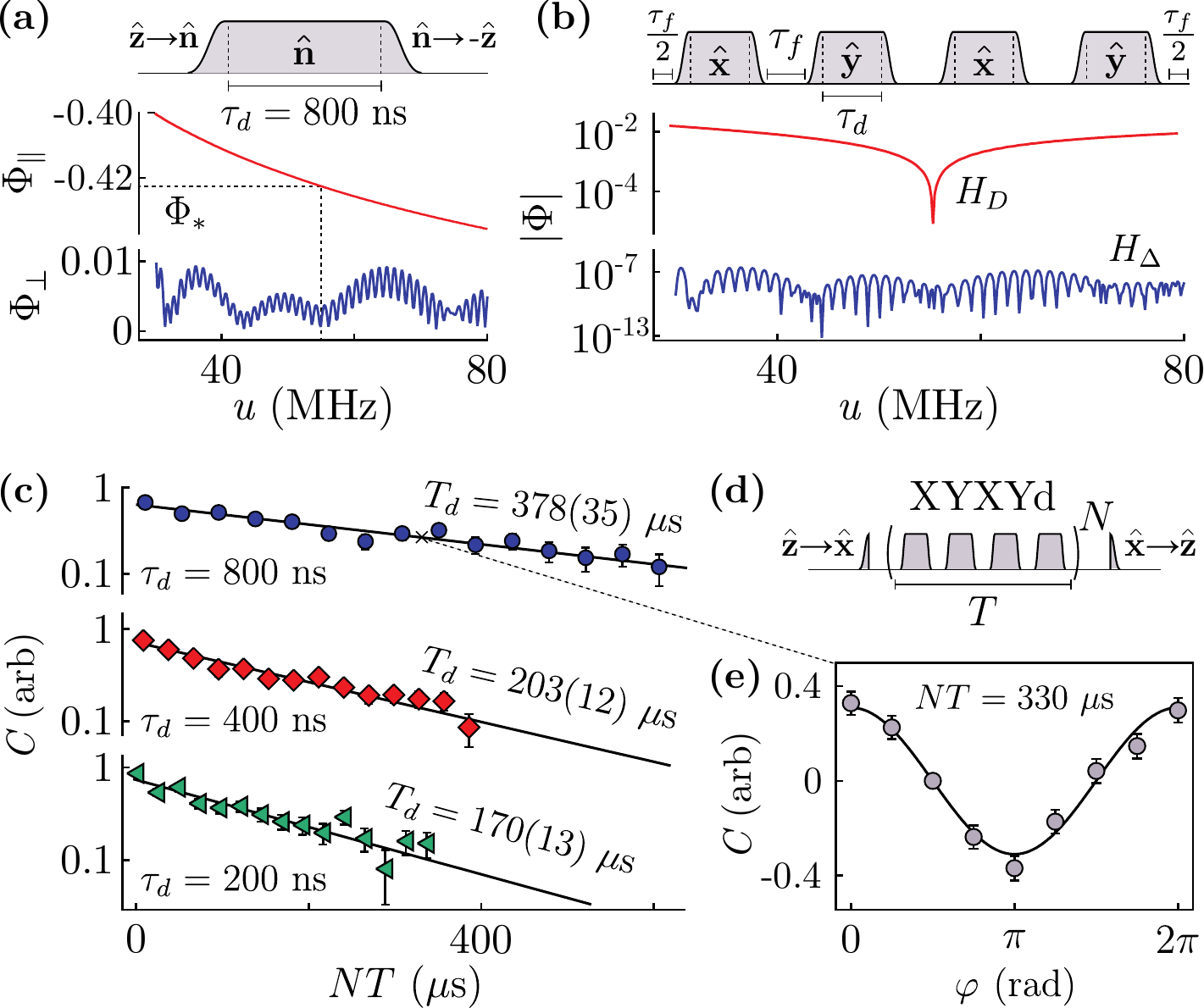}
    \caption{
        XYXYd sequence design and simulation.
        (a) Numerical evaluation of the dipolar rescaling factors 
        $\Phi_{\parallel}(H_D)$ (red) and $\Phi_{\perp}(H_D)$ (blue) for a single 
        $\uvec{n}$-primitive with $\tau_d = 800$~ns, showing a rescaling of 
        $\Phi_* = -0.42$ at the center of the Rabi distribution ($u = 55$~MHz). Pulses labeled as $\uvec{n}_1\rightarrow\uvec{n}_2$ denote adiabatic half passages (AHPs) that map $\uvec{n}_1$ to $\uvec{n}_2$ on the Bloch sphere.
        The AHPs were 67~ns long, and designed using the 
        optimal control protocol given in Ref.~\cite{Tabatabaei2021}.
        (b) Dipolar $|\Phi(H_D)|$ and resonance offset $|\Phi(H_\Delta)|$ suppression metrics for a single XYXYd block.
        (c) Measured XYXYd decay curves and associated exponential fits (solid lines) as a function of the total XYXYd time $NT$. 
        (d) Pulse sequence used in (c), consisting of $N$ repetitions of an XYXYd block of duration $T$.
        (e) Phase-accumulation test used to rule out spin-locking. 
        An effective $z$-field is introduced by applying a phase shift 
        of $+\varphi/(8N)$ to each $\ux$-primitive and 
        $-\varphi/(8N)$ to each $\uy$-primitive. 
        The observed modulation in the signal as a function of $\varphi$ rules out spin-locking during the XYXYd sequence.
    }\label{fig2}
\end{figure}

To suppress electron-electron dipole interactions, we replace each $\pi$ pulse by a composite adiabatic inversion consisting of a resonant drive of duration $\tau_d$, sandwiched between two short adiabatic half passages (AHPs) [Fig.~\ref{fig2}(a)].
We refer to this composite inversion as the \textit{$\uvec{n}$-primitive},
where $\uvec{n}$ is the constant transverse control axis set by the microwave phase
[$\mathbf{a}(t)=\uvec{n}$ during the driven segment.]

In the limit of large Rabi frequencies, $2\pi u(\rr_j ) \gg \vert D_{jk}\vert$, the rotating frame unitary describing the central drive segment is
\begin{align}
    U_d(\uvec{n}) \simeq e^{-i2\pi\tau_d \sum_j u(\rr_j) \uvec{n}\cdot\mathbf{S}_j} e^{-i\tau_d \bar{H}_d(\uvec{n})},
\end{align}
where
\begin{align}
    \bar{H}_d(\uvec{n}) = -\frac{1}{2} \sum_{j<k} D_{jk} [3( \uvec{n}\cdot\mathbf{S}_j)(\uvec{n}\cdot\mathbf{S}_k) - \mathbf{S}_j \cdot \mathbf{S}_k],
\end{align}
is the effective Hamiltonian in the interaction frame of $H_C(t)$, i.e. the toggling frame \cite{haeberlen1976}, and $\tau_d$ is the drive duration.
Hence, for ideal AHPs that are much faster than the time scale of $H_D$ and $H_\Delta$, the $\uvec{n}$-primitive propagator is
\begin{align}
    U(\uvec{n}) &= U_{\uvec{n}\rightarrow-\uz}~U_d(\uvec{n})~U_{\uz\rightarrow\uvec{n}} \nonumber\\[5pt]
    &= U_{\uz\rightarrow-\uz}~e^{-i2\pi\tau_d \sum_j u(\rr_j) S_j^z}~e^{-i\tau_d \bar{H}_d(\uz)},
\end{align}
where we use the notation $U_{\uvec{n}_1\rightarrow\uvec{n}_2}$ to denote a generic unitary that maps the point $\uvec{n}_1$ to $\uvec{n}_2$ on the Bloch sphere.
Since $\bar{H}_d(\uz) = - H_D/2$, an ideal $\uvec{n}$-primitive results in a combination of (1) time-reversed dipolar evolution with half the coupling, (2) a Rabi-frequency-dependent rotation around $\uz$ and (3) a $\pi$ rotation.
Therefore, for ideal AHPs, an XYXYd sequence [Fig.~\ref{fig2}(b)] with free-evolution times $\tau_f = \tau_d/2$ would average out $H_D$ and $H_\Delta$ to leading order:

\begin{align}
     \left[U_0(\tau_f/2) U(\uy) U_0(\tau_f)U(\ux) U_0(\tau_f/2)\right]^2 \propto \id,
\end{align}
with $U_0(\tau) \equiv e^{-i\tau (H_D+H_\Delta)}$ being the free-evolution propagator.

In practice, dipolar evolution during the AHPs is non-negligible.
To quantify the net effect of a control block on an operator $h$, we numerically evaluate its zeroth-order average Hamiltonian $\bar{h}^{(0)}$ in the toggling frame, and decompose it into components parallel and orthogonal to $h$ with respect to the Hilbert-Schmidt inner product:
\begin{align}
    &\Phi_\parallel(h) =\langle h, \bar{h}^{(0)}\rangle/\norm{h}^2,\\
    &\Phi_\perp(h) = \norm{\bar{h}^{(0)}-\Phi_\parallel(h) h }/\norm{h},
\end{align}
where $\Phi_\parallel(h)$ quantifies the retained fraction of $h$, and $\Phi_\perp(h)$ measures leakage into orthogonal directions.
We use $\abs{\Phi(h)} = \sqrt{\Phi_\parallel^2(h) + \Phi_\perp^2(h)}$ as a single metric for the suppression of $h$ by the sequence.

For the numerical evaluation of $\bar{h}^{(0)}$, we consider a two-spin system at fixed Rabi frequency $u$.
Figure~\ref{fig2}(a) shows $\Phi_\parallel(H_D)$ and $\Phi_\perp(H_D)$ for a single $\tau_d = 800$~ns $\uvec{n}$-primitive, showing a dipolar rescaling factor of $\Phi_* = -0.42$ at the center of the Rabi distribution, with less than $1\%$ leakage into the orthogonal subspace. 
Accordingly, choosing $\tau_f = -\Phi_* (\tau_d + 2\tau_{\mathrm{AHP}})$, where $\tau_{\mathrm{AHP}} = 67$~ns is the AHP duration, results in an XYXYd sequence with $\abs{\Phi(H_D)} < 10^{-2}$ across the Rabi distribution [Fig.~\ref{fig2}(b)].
Repeating the analysis for the resonance offset term gives $\abs{\Phi(H_\Delta)} \leq 10^{-7}$.
Because of the dipolar-averaging condition, $\tau_d$, $\tau_f$, and $T$ are not independent; specifying the total XYXYd block duration $T$ uniquely fixes $\tau_d$ and $\tau_f$.

Using this approach, we designed XYXYd sequences of varying drive durations and measured the resulting coherence times.
The data for three representative drive durations are shown in Fig.~\ref{fig2}(c), with the complete set of extracted time constants provided in Table~\ref{tab:table1}.
For $\tau_d = 800$~ns, we measured a coherence time of $T_d = 378(35)~\mu$s.
We ruled out spin-locking effects by measuring oscillations generated by an effective $z$-field introduced via controlled phase shifts in the sequence [Fig.~\ref{fig2}(e)] \cite{Boutis2003}.

We note that the XYXYd sequence can also be used for Fourier imaging by
applying static field gradients during the $\tau_f$ intervals, in which case the relevant coherence time, $T_{d,\mathrm{free}}$, is in units of accumulated free evolution time (Table~\ref{tab:table1}).
For $\tau_d = 800$~ns, we measure $T_{d,\mathrm{free}} = 112(10)~\mu\mathrm{s}$---a factor of $19$ longer than the bare $T_2$.

\begin{table}[h!]
\caption{
Sequence parameters and measured XYXYd decay constants in terms of the total ($T_d$) and free evolution ($T_{d,\mathrm{free}})$ times.  \label{tab:table1}
}
\begin{ruledtabular}
\begin{tabular}{ccc cc}
\multicolumn{3}{c }{{Sequence parameters}} &
\multicolumn{2}{c}{{Decay times}} \\[3pt]
$\tau_d$ (ns) & $\tau_f$ (ns) & $T~(\mu\text{s})$ &
$T_d~(\mu\text{s})$ & $T_{d,\text{free}}~(\mu\text{s})$ \\[2pt]
\hline\\[-8pt]
200 & 95 & 1.717 & $170(13)$ & $38(3)$ \\
400 & 192 & 2.903 & $203(12)$ & $54(3)$ \\
565 & 278 & 3.913 & $151(16)$ & $43(5)$ \\
800 & 395 & 5.317 & $378(35)$ & $112(10)$ \\
1311 & 560 & 7.303 & $457(48)$ & $140(15)$ \\
1600 & 795 & 10.117 & $478(49)$ & $150(15)$ \\
\end{tabular}
\end{ruledtabular}
\end{table}

%%%%%%%%%%%%%%%%%%%%% External Field Sensing %%%%%%%%%%%%%%%%%%%%%%%%%%%%%%
\section{External-Field Sensing}\label{sec:extField}

\begin{figure*}[p!]
    \centering
    \includegraphics[width=1\linewidth]{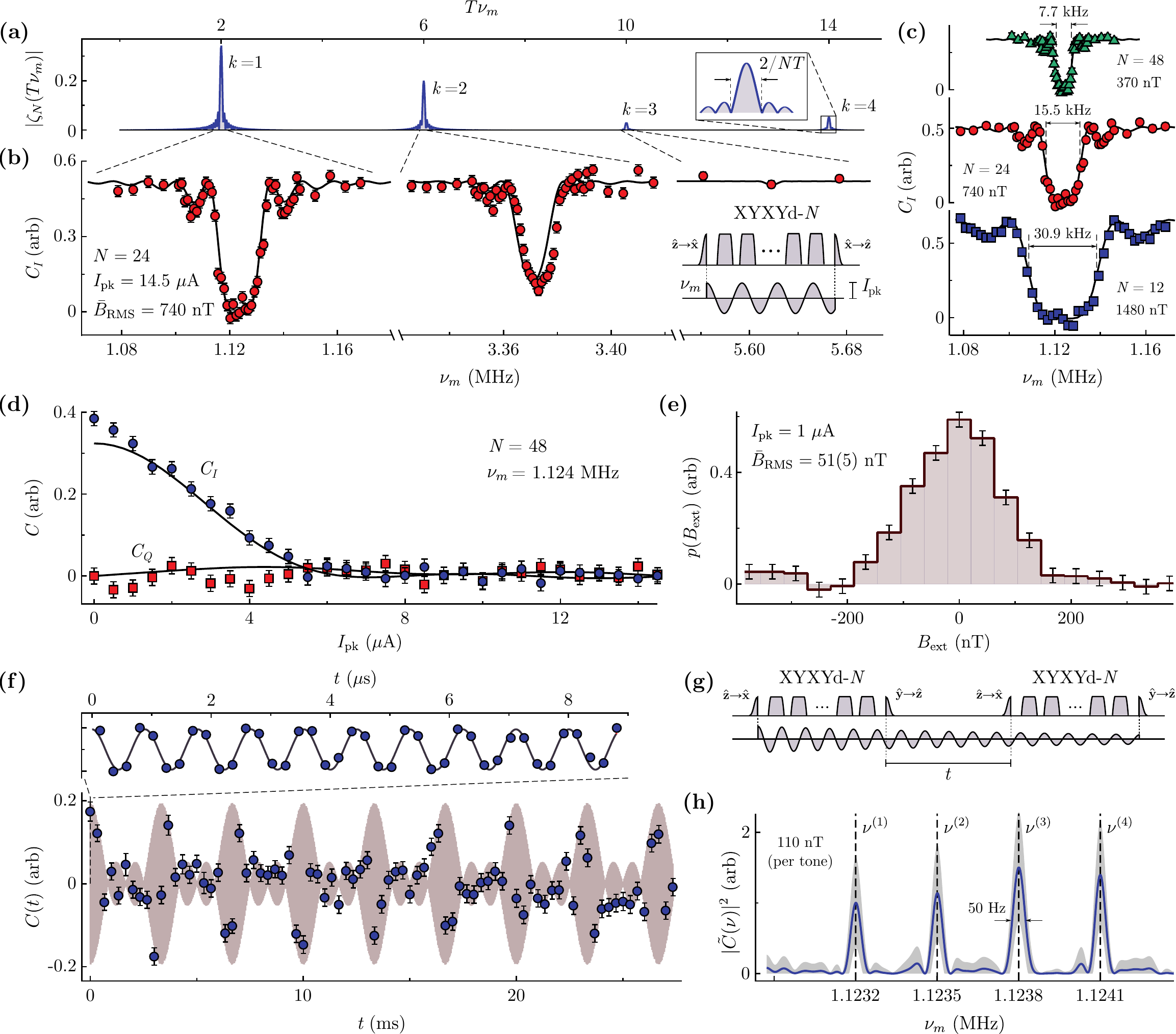}
    \caption{Quantum sensing of external fields.
    (a) XYXYd-$N$ filter function magnitude as a function of normalized drive frequency $T\nu_m$, showing the first four filter function lobes.
    (b) Measured in-phase signal $C_I$ as a function of the drive frequency $\nu_m$ for $N=24$ repetitions of the XYXYd block ($T=1.780~\mu$s).
    The modulation was applied by passing a peak current of $I_\mathrm{pk} = 14.5~\mu$A through the CFFGS during the XYXYd sequence (inset).
    The corresponding ensemble-averaged RMS field, inferred from the distribution in (e), was 740(68)~nT.
    (c) Measured in-phase signal for the $k=1$ lobe at different XYXYd repetitions.
    The $N\times I_\mathrm{pk}$ product was kept constant to keep the phase accrued by the electron the same at the center of the lobe.
    (d) In-phase and quadrature signals measured at the center of the first lobe as a function of the peak CFFGS current $I_\mathrm{pk}$.
    The quadrature signal was measured by changing the phase of the last AHP by $\pi/2$.
    (e) Longitudinal-field distribution $p(B_\mathrm{ext})$ of the ensemble for $1~\mu$A peak current, extracted from the Fourier transform of the data in (d). 
    The distribution has a full width at half maximum of $210$~nT, reflecting the spatial inhomogeneity of the CFFGS field across the detected ensemble.
    All solid lines in (b-d) are calculations using Eq.~(\ref{eq:CICQ}) and the measured $B_\mathrm{ext}$ distribution (e).
    (f) Time-domain correlation-spectroscopy signal, (bottom) with and (top) without undersampling.
    The shaded curve represents the expected modulation $\propto \sum_{n=1}^4 \cos[2\pi \nu^{(n)} t]$.
    (g) The correlation spectroscopy pulse sequence.
    For the measurements in (f), we used $N=24$.
    (h) Power spectral density (PSD) of the undersampled data in (f), showing four peaks at the expected frequencies $\nu^{(n)}$ of the applied field.
    Zero-padding was used for visualization of the PSD.
    All measurements are taken with a $T = 1.780~\mu$s XYXYd duration.
    The shaded areas correspond to 68\% confidence regions.
    All reported RMS fields have a common relative uncertainty of $9\%$.
    }
    \label{fig3}
\end{figure*}

To demonstrate AC magnetic-field sensing with SQUINT, we apply the XYXYd sequence to detect an RF magnetic field generated by the CFFGS.
The basic idea is analogous to AC magnetometry protocols used in NV platforms \cite{Taylor2008}.
By timing the XYXYd applications relative to the oscillation of the applied field, each electron spin accrues a coherent phase proportional to the longitudinal component of that field.
The evolution under $N$ applications of the XYXYd sequence (denoted XYXYd-$N$) is governed by the effective Hamiltonian $H_\mathrm{eff} = -\sum_j \Omega_N(\rr_j,T)\, S_j^z$,
where
\begin{align}
    \Omega_N(\rr,T) \equiv \gamma_e \int_{-\infty}^\infty d\nu~\zeta_N(T\nu) \tilde{B}_z(\rr,\nu) e^{i\pi N T\nu}, \label{eq:effResOff}
\end{align}
and $\tilde{B}_z(\rr,\nu)= \int_{-\infty}^\infty dt ~B_z(\rr,t)e^{-i2\pi \nu t}$ is the Fourier transform of the longitudinal field $B_z(\rr,t)$.
The filter function
\begin{align}
    \zeta_N(T\nu) \equiv \sum_{k=-\infty}^\infty \Gamma_k~\sinc\left[NT \left(\nu-\nu_k\right)\right],
\end{align}
is the sum over sinc function lobes centered at $\nu_k = 2(2k-1)/T$, with weights $\Gamma_k$ given by the Fourier components of the toggling-frame $S_j^z$ modulation (see Supplemental Material \cite{Supplements}).
These weights determine the sensitivity of the sequence to field components at each harmonic.
Here, $\sinc(x) \equiv \sin(\pi x)/(\pi x)$ denotes the normalized sinc function.
Hence, the spectral components of $B_z$ that overlap with the filter function result in a measurable modulation of the transverse magnetization, enabling frequency-selective sensing of the applied field.

It is useful to distinguish the response of XYXYd-$N$ from conventional Carr-Purcell or XYXY magnetometry implemented with hard pulses.
In the hard-pulse limit, control pulses are treated as instantaneous, and the sensed phase accumulates only during free evolution.
This approximation is not valid for XYXYd, where a substantial fraction of each block is spent under microwave drive to average dipolar interactions.
We therefore compute the accumulated phase from the full time dependence of the sequence, encoded in the filter function $\zeta_N$ in Eq.~(\ref{eq:effResOff}).
For an XYXYd-$N$ sequence applied at $t=0$, an electron spin subject to $B_z(t)=B_\mathrm{ext}\cos(2\pi\nu_m t)$ accumulates $\varphi(B_\mathrm{ext}) = NT\Omega_N = NT\gamma_e B_\mathrm{ext}\cos(\pi NT\nu_m)\zeta_N(T\nu_m)$.
At the center of the $k^\text{th}$ filter-function lobe, this becomes $\varphi(B_\mathrm{ext}) = NT\gamma_e B_\mathrm{ext}\Gamma_k$.

Operating at a larger index $k$ allows the sequence to detect higher frequencies without reducing the XYXYd block duration $T$.
However, the Fourier weights $\Gamma_k$ become smaller as $k$ increases, thus reducing the detection sensitivity.
In practice, the optimal sensitivity therefore reflects a tradeoff between shortening $T$ and operating at a higher $k$.
The calculation of $\Gamma_k$, along with the values used in this work, is provided in Section VIII of the Supplemental Material \cite{Supplements}.

Figure~\ref{fig3} summarizes the external-field sensing measurements.
By driving the CFFGS at frequency $\nu_m$ with peak current $I_\mathrm{pk}$, we generated a spatially non-uniform RF field across the sensor ensemble, resulting in each spin accruing the phase $\varphi(B_\mathrm{ext})$ according to its local field amplitude.
The resulting modulation in the transverse magnetization components can be converted to an observable correlation signal using additional AHPs [Fig.~\ref{fig3}(b) inset].
The corresponding signal quadratures are
\begin{align}
    \begin{bmatrix}C_I\\C_Q\end{bmatrix} \propto \int_{-\infty}^\infty dB_{\mathrm{ext}}~p(B_{\mathrm{ext}}) \begin{bmatrix}\cos\varphi(B_{\mathrm{ext}})\\
    \sin\varphi(B_{\mathrm{ext}})\end{bmatrix},\label{eq:CICQ}
\end{align}
where $p(B_{\mathrm{ext}})$ is the readout-gradient-weighted distribution of the $z$-field amplitude over the sample.

Figures~\ref{fig3}(b,c) show the measured in-phase signal $C_I$ for a $T = 1.780~\mu$s XYXYd-$N$ sequence, as a function of the drive frequency $\nu_m$ for the first three filter function lobes and several values of $N$.
As $N$ increases, the spectral dips narrow, reflecting the narrowing of the filter function lobes, and thus enhanced frequency selectivity.
For $N=48$, we measured a full width at half maximum of $7.7$~kHz for the spectral dip.
The maximum usable $N$ is ultimately limited by decoherence under XYXYd.
For the sequence used here, we measured a coherence time $T_d = 104(5)~\mu$s, corresponding to $58(3)$ XYXYd repetitions.

In a separate measurement, we fixed the drive frequency at the center of the first lobe $\nu_m=2/T=1.124~\mathrm{MHz}$, and measured $C_I$ and $C_Q$ as a function of the peak current $I_\mathrm{pk}$ [Fig.~\ref{fig3}(d)] with $\zeta_N(T\nu_m) = \Gamma_1=0.34$, and $\varphi(B_\mathrm{ext}) =  NT \gamma_e \Gamma_1  B_\mathrm{ext} $.
We introduce the parameter $\kappa \equiv NT\gamma_e \Gamma_1 I_{\mathrm{pk}}/I_{\mathrm{ref}}$, where $I_{\mathrm{ref}} = 1~\mu$A is a fixed reference current.
Writing Eq.~(\ref{eq:CICQ}) in complex form gives $C_I+iC_Q \propto \int dB_\mathrm{ext}~p(B_\mathrm{ext}) e^{i\kappa B_\mathrm{ext}}$, such that a Fourier transform over $\kappa$ yields the distribution $p(B_\mathrm{ext})$ associated with the reference current [Fig.~\ref{fig3}(e)].
The sample-averaged RMS field is $\bar{B}_\mathrm{RMS} = \int dB_\mathrm{ext}~\abs{B_\mathrm{ext}}~p(B_\mathrm{ext})/\sqrt{2} = 51(5)~$nT.
Modeling the measurements in Fig.~\ref{fig3}(b--e) using Eq.~(\ref{eq:CICQ}) and the extracted $p(B_\mathrm{ext})$ shows good agreement with the data (solid lines).

The spectral resolution provided by the XYXYd filter function is ultimately limited by the electron coherence time $T_d$.
To enable higher resolution, we employ a correlation spectroscopy sequence \cite{Jiang2023, Laraoui2013} constructed from two XYXYd-$N$ blocks separated by a free evolution time $t$, during which the accrued phase from the first block is stored in the longitudinal spin components [Fig.~\ref{fig3}(g)].
In the limit where the electron accrues a phase $\ll 1$ during each XYXYd-$N$ block, a measurement of the signal after the second block yields the correlation between the electron phases accrued during the two blocks.
The Fourier transform of this signal results in a high-resolution spectrum of the applied field.
Theoretical details of the correlation-spectroscopy experiments are provided in the Supplemental Material \cite{Supplements}.

To demonstrate this method, we applied a four-tone current $I(t) = I_\mathrm{pk}\sum_{n=1}^4 \cos[2\pi \nu^{(n)} t]$ to the CFFGS, with $I_\mathrm{pk}=2.2~\mu$A per tone and frequencies $\nu^{(n)}$ chosen within the first filter-function lobe, separated by 300~Hz.
The drive amplitude for each tone corresponds to a sample-averaged RMS $z$-field of 110(10)~nT.
The resulting time-domain signal is shown in Fig.~\ref{fig3}(f).
Rather than sampling the oscillations at the Nyquist rate, we undersampled the signal at 3~kHz sampling rate, chosen such that the frequency band of interest $\mathrm{[1.12287~MHz,~1.12437~MHz]}$ aliases without overlap into the first Nyquist zone $\mathrm{[0,~1.5~kHz]}$, enabling recovery of the spectrum using fewer measurements.

By Fourier transforming the undersampled data, we constructed the power spectral density (PSD) of the four-tone modulation [Fig.~\ref{fig3}(h)].
We note that the longest evolution time used in this measurement was $\orderof 30$ ms, which is well below the measured electron spin-lattice relaxation time [$T_1 = 671(24)$~ms] and therefore did not approach the $T_1$-limited maximum time available for correlation measurements. In principle, extending the correlation delay toward the $T_1$ limit would enable spectral resolutions on the order of a few hertz.

%%%%%%%%%%%%%%%%%%%%% Nuclear Spin Sensing %%%%%%%%%%%%%%%%%%%%%%%%%%%%%%

\section{Nuclear Spin Sensing and Spectroscopy}

We use the XYXYd sequence to probe the local nuclear spins surrounding each electron-spin sensor molecule.
Hyperfine coupling to nearby nuclei produces a narrowband, stochastic magnetic field at the nuclear Larmor frequency that contributes to the longitudinal field experienced by the electron.
The sensing principle mirrors the external-field case: by tuning the XYXYd filter function to the Larmor frequency of a given nuclear species, the electron spin accumulates phase from its coupling to those nuclei, resulting in a measurable suppression of the spin echo.

\begin{figure*}[p!]
    \centering
    \includegraphics[width=\linewidth]{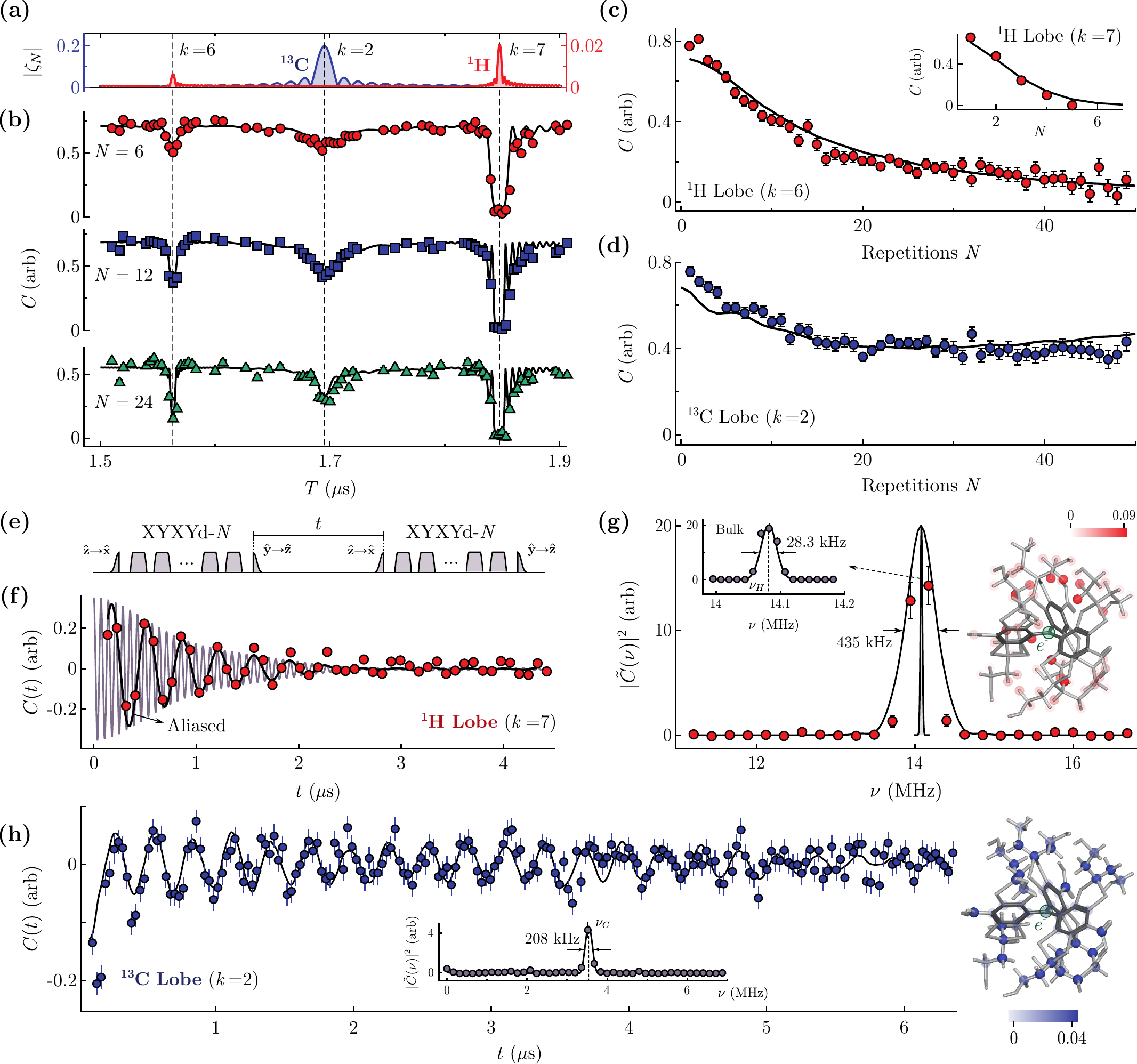}
    \caption{Nuclear spin sensing and spectroscopy using the XYXYd sequence.
    (a) Magnitude of the XYXYd filter function $\zeta_N(T\nu_{\mathrm{nuc}})$ as a function of $T$, with $\nu_{\mathrm{nuc}}$ evaluated at the ${}^1$H and ${}^{13}$C Larmor frequencies.
    The two lobes at $T = 1.563~\mu$s and $T = 1.847~\mu$s correspond to the $k=6,7$ harmonics of the ${}^{1}$H filter function, while the lobe at $T = 1.693~\mu$s corresponds to the $k=2$ harmonic for ${}^{13}$C.
    The separate vertical axes correspond to the ${}^1$H (red) and ${}^{13}$C (blue) nuclei.
    (b) Nuclear spin sensing measurements obtained by sweeping the XYXYd duration $T$ for different numbers of repetitions $N$.
    Echo suppression is observed when a filter-function harmonic is tuned to the ${}^1$H or ${}^{13}$C Larmor frequency.
    Solid lines in all panels denote simulations based on intramolecular ${}^1$H and ${}^{13}$C nuclei, unless stated otherwise.
    (c,d) Correlation signal measured with the XYXYd filter function placed at the ${}^1$H and ${}^{13}$C resonances, respectively, as a function of the number of XYXYd repetitions $N$.
    (e) Correlation spectroscopy pulse sequence, consisting of two XYXYd-$N$ blocks separated by a variable free-evolution time $t$.
    (f) Time-domain correlation signal measured at the ${}^1$H resonance. 
    Similar to the external field sensing experiments, the data was intentionally undersampled to reduce the number of measurements.
    (g) Frequency spectrum of the ${}^1$H correlation signal reveals a hyperfine-broadened local lineshape relative to the independently measured bulk proton NMR spectrum (left inset).
    The solid curve in the inset depicts a Gaussian fit to the bulk proton lineshape.
    (h) Time-domain correlation signal for the ${}^{13}$C resonance, along with the corresponding frequency spectrum (lower inset).
    Right insets of (g) and (h): Spatial maps of the fractional contribution of each ${}^1$H and ${}^{13}$C nucleus on the OX063 molecule to the $t=0$ correlation signal, respectively.
    The total signal is normalized to 100\%.
    All quoted spectral widths correspond to full widths at half maximum.
    Acquisition times for these measurements are provided in Section~XI of the Supplemental Material \cite{Supplements}.
    }
    \label{fig4}
\end{figure*}

Throughout this section, we model all nuclei as $I=1/2$ spins and, because XYXYd effectively decouples the electron spins from one another, we analyze the dynamics on a single-electron basis.
In the high-field limit, the hyperfine interaction for a given electron spin is $H_\mathrm{HF}
= \sum_{l} \left( A_{0l} I_l^z + \mathbf{A}_{1l}\cdot\mathbf{I}_l \right) S^z$, where $A_{0l}$ is the secular coupling to the $l^\text{th}$ nucleus, and the transverse vector $\mathbf{A}_{1l}$ represents the pseudo-secular coupling.
For an axially symmetric hyperfine tensor, these couplings depend on the angle $\theta$ between the tensor's non-degenerate principal axis and the static field.
The isotropic-anisotropic decomposition of the hyperfine tensor gives \cite{Schweiger2001}
\begin{align}
A_0 &= A_\mathrm{iso}
+ A_\mathrm{aniso}\left(1-3\cos^2\theta\right), \label{eq:A0}\\
A_1 &\equiv \abs{\mathbf{A}_1} =\frac{3}{2} A_\mathrm{aniso} \sin(2\theta). \label{eq:A1}
\end{align}

Under an XYXYd-$N$ sequence, the evolution is described by the effective Hamiltonian $H_\mathrm{eff}= \sum_{l}\zeta_N(T\nu_{\mathrm{nuc}})  A_{1l} I_l^x S^z$ in the interaction frame of the nuclear Zeeman term (Supplemental Material \cite{Supplements}).
Here, $\nu_{\mathrm{nuc}}$ is the nuclear Larmor frequency.
For a given realization of the nuclear spin bath, evolution under $H_\mathrm{eff}$ results in a phase $\varphi = N T \sum_{l} \zeta_N(T\nu_{\mathrm{nuc}})\, A_{1l} I_l^x$ accrued by the electron.
As in the external-field measurements, this phase is converted into a measurable signal by applying AHPs before and after the XYXYd-$N$ sequence, such that the contribution to the signal by each electron is
\begin{align}
     C(t) \propto \langle \cos\varphi\rangle_I = \prod_l \cos\left(\frac{NT \zeta_N(T\nu_{\mathrm{nuc}}) A_{1l} }{2}\right), \label{eq:cosphi}
\end{align}
leading to a suppression of the electron spin echo when the filter function is tuned to the nuclear Larmor frequency.
Here, $\langle \cdot\rangle_I$ denotes averaging over nuclear spin configurations.
The measured signal is the sum of single-electron contributions given by Eq.~(\ref{eq:cosphi}).

Figure~\ref{fig4}(b) shows nuclear spin sensing measurements for different values of $N$, obtained by sweeping the XYXYd duration $T$, and hence the location of the filter function lobes.
We observe echo suppression at three values of $T$, with two arising from different harmonics of the filter function at the ${}^1$H Larmor frequency $\nu_H = 14.08$~MHz and a third corresponding to ${}^{13}$C nuclei with $\nu_C = 3.54$~MHz [Fig.~\ref{fig4}(a,b)].
The higher Larmor frequency of ${}^1$H necessitated detection at higher harmonics of the filter function, reducing sensitivity. This loss is not intrinsic to the sensing scheme and could be mitigated by operating at a lower static field, thereby lowering $\nu_H$ and permitting detection with lower filter harmonics.

We further characterized the echo suppression by positioning the XYXYd filter function at a particular Larmor frequency, and varying the number of XYXYd repetitions $N$.
Fig.~\ref{fig4}(c,d) show the resulting correlation signal.
We independently measured the decay expected from the finite electron coherence time under XYXYd and divided the measured signal by this calibrated decay to remove its contribution.
For both spin species, increasing $N$ leads to a progressive suppression of the echo.
For the ${}^{13}$C resonance, the emergence of a non-zero plateau at large $N$ indicates that a fraction of OX063 molecules contain an electron spin with negligible coupling to any ${}^{13}$C nuclear spin.
The ratio of the initial signal to the large-$N$ plateau yields an estimated fraction of $0.49(2)$ of OX063 molecules whose electron spin is appreciably coupled to at least one ${}^{13}\mathrm{C}$ spin.
Given the ensemble size, this corresponds to around 70 OX063 molecules contributing to the observed contrast.

To model these measurements, we calculate the expected echo suppression using only the ${}^1$H and ${}^{13}$C nuclei on the OX063 molecule, while averaging over molecular orientations and the electron Rabi frequency distribution [Fig.~\ref{fig1}(e)].
For ${}^1$H nuclei, we use Eq.~(\ref{eq:cosphi}) for the 63 protons on the molecule, with hyperfine couplings computed from Eq.~(\ref{eq:A0},\ref{eq:A1}) using the point-dipole approximation: $A_{\mathrm{iso}} = 0$ and $A_{\mathrm{aniso}} = \mu_0 \hbar \gamma_H\gamma_e/\left(4\pi r^3\right)$, where $\gamma_H/(2\pi) = 42.6$~MHz/T is the ${}^1$H gyromagnetic ratio and $r$ is the nucleus-electron distance.
The ${}^{13}$C nuclei require a different treatment because, at natural abundance, each of the 52 carbons only has a $1\%$ probability of being ${}^{13}$C, and several carbon sites lie sufficiently close to the electron that their strong hyperfine couplings fall outside the regime where the effective-Hamiltonian description applies.
We therefore model the ${}^{13}$C response using a simulation of the time-dependent Hamiltonian under the applied control waveform, with hyperfine parameters taken from the literature \cite{Trukhan2013,Bowman2005}.
Details of the simulations are provided in the Supplemental Material \cite{Supplements}.

The simulations [solid lines in Fig.~\ref{fig4}(b-d)] reproduce the experimental data, indicating that the signal is dominated by nuclei on the OX063 molecule itself rather than by the surrounding matrix.
This conclusion is further supported by the experimentally extracted fraction, $0.49(2)$, of molecules exhibiting appreciable ${}^{13}\mathrm{C}$ hyperfine coupling, consistent with the expected ${}^{13}\mathrm{C}$ occupancy in OX063 at natural abundance.

As in the external-field case, the XYXYd sequence can be extended to correlation spectroscopy using the sequence given in Fig.~\ref{fig4}(e).
The filter function is positioned at the selected nuclear Larmor frequency.
The measured signal $C(t)$ probes the correlation between the electron phases accumulated in the two XYXYd-$N$ blocks, which is modulated by nuclear spin dynamics under secular hyperfine couplings during the correlation delay $t$.
In the small-phase regime, the signal is approximately
\begin{align}
    C(t) \propto  \sum_l N^2T^2\Gamma^2 A_{1l}^2 \cos\left(\frac{ t A_{0l}}{2}\right) \cos(2\pi\nu_{\mathrm{nuc}} t).\label{eq:corrSpec}
\end{align}
Here, $\Gamma$ is the filter function weight.
For a derivation of Eq.~(\ref{eq:corrSpec}), see the Supplemental Material \cite{Supplements}. 

Figures~\ref{fig4}(f,h) show the resulting time-domain correlation signals for the ${}^1$H and $^{13}$C resonances, respectively.
The corresponding power spectra are shown in Fig.~\ref{fig4}(g) and the inset of Fig.~\ref{fig4}(h).
Simulations based on the effective Hamiltonian for ${}^1$H nuclei and time-dependent waveform simulations for ${}^{13}$C nuclei, using the same modeling assumptions as before are in good agreement with the measured spectra.
The broadening of the local ${}^{1}$H spectrum relative to the independently measured bulk lineshape [Fig.~\ref{fig4}(g) left inset] directly indicates that the detected nuclei are local to the sensor spins.
We note that the sensing protocol is directly compatible with electron- or nuclear-spin decoupling sequences applied during the correlation delay to suppress the hyperfine evolution and thereby narrow the lineshape.

We compute the fractional contribution to the correlation signal at $t=0$ for each nuclear site on the molecule (see Supplemental Material \cite{Supplements} for details).
The resulting spatial maps for both nuclei are shown in the insets of Fig.~\ref{fig4}(g,h).
The ${}^1$H signal is distributed across the molecule with an $A_{1}^2 \propto 1/r^6$ weighting [Eq.~(\ref{eq:corrSpec})]; in particular, 90\% of the signal arises from 38 of the 63 hydrogen sites in OX063.
For ${}^{13}\mathrm{C}$, the 19 sites comprising the central carbon and the three phenyl rings contribute negligibly because their hyperfine couplings are comparable to, or exceed, the electron Rabi frequency; instead, 90\% of the signal arises from 26 of the remaining sites.
Given the 1\% natural abundance, this corresponds to an expected number of about 36 ${}^{13}$C nuclei contributing 90\% of the detected signal within the measured sample volume.
These measurements demonstrate frequency-selective detection and spectroscopy of a small, local ensemble of nuclear spins using mechanically detected molecular spin sensors.

%%%%%%%%%%%%%%%%%%%%% Sensitivity and Outlook %%%%%%%%%%%%%%%%%%%%%%%%%%%%%%
\section{Sensitivity and Outlook}~\label{sec:sensitivity}

In this section, we quantify the sensitivity of the SQUINT platform and outline potential improvements for future applications.
In all cases considered here, we determine sensitivity by comparing the electron-spin signal produced by the quantity of interest---either an external magnetic field or coupling to a nuclear spin---to the noise expected for a given acquisition time.
The signal is set by the phase accumulated by the electron spins under the XYXYd sequence, whereas the noise arises from the thermal force noise of the mechanical oscillator and fluctuations of the sensor-spin ensemble (spin noise).

For the statistical correlation measurements performed using the MAGGIC spin detection protocol, the signal-to-noise ratio (SNR) has been analyzed in detail in Ref.~\cite{Tabatabaei2024}.
A summary of the necessary equations is also provided in the Supplemental Material \cite{Supplements}.
For a single spin experiencing a local readout gradient $G$, we define the force-noise equivalent bandwidth $\mathcal R\equiv \mu^2D^2G^2/(2S_F)$.
Physically, $\mathcal R$ is the bandwidth in which the oscillator thermal-force noise has integrated variance equal to the force variance of that spin.
Larger $\mathcal R$ therefore corresponds to improved mechanical readout.

For an ensemble of $N_s$ sensor spins, the standard deviation of a single shot of the correlation measurement may be written in the form $\sigma(N_s\mathcal{R},\tau_0,\tau_m)$, where $\mathcal{R}$ is interpreted as an ensemble-averaged single-spin bandwidth, obtained by replacing $G^2$ with its average $\overline{G^2}$.
Here, $\tau_0$ is the measure block duration [Fig.~\ref{fig1}(d)], $\mu$ is the electron magnetic moment, $S_F$ is the (single-sided) thermal force noise spectral density of the nanomechanical oscillator, and $D$ and $\tau_m$ are the duty cycle and spin correlation time under MAGGIC, respectively \cite{Tabatabaei2024, Haas2022, Rose2018}.
Throughout this section, we normalize the expectation value of the total electron signal in the absence of a control sequence to unity, such that
$1/\sigma(N_s\mathcal{R},\tau_0,\tau_m)$
is the single-shot SNR of the bare signal.

For sensing external fields, we consider an XYXYd-$N$ sequence similar to the inset of Fig.~\ref{fig3}(b), with the phase of the last AHP shifted by $\pi/2$.
For a sinusoidal longitudinal field with amplitude $B_\mathrm{ext}$ centered at a filter-function lobe with weight $\Gamma$, the resulting correlation signal takes the form $\sin(NT\Gamma \gamma_e B_\mathrm{ext}) e^{-NT/T_d}$.
For a fixed total acquisition time $\tau_{\mathrm{acq}}$, the SNR is
\begin{align}
\mathrm{SNR}(B_\mathrm{ext},\tau_0,\tau_m,N)
&=
\sqrt{\frac{\tau_\mathrm{acq}}{NT + \tau_0 + \tau_\mathrm{ovh}}}
 \times \nonumber\\[5pt]
 & \hspace{12pt} \frac{\sin(NT\Gamma \gamma_e B_\mathrm{ext})\, e^{-NT/T_d}}{\sigma(N_s\mathcal{R}, \tau_0,\tau_m)},
\end{align}
where $\tau_\mathrm{ovh} = 2~\mathrm{ms}$ accounts for the overhead per shot arising from finite rise and fall times in the MAGGIC protocol \cite{Rose2018}.
The detection threshold $B_{\min}(N_s\mathcal{R},\tau_\mathrm{acq})$ is defined as the smallest field amplitude for which $\mathrm{SNR}(B_{\min},\tau_0,\tau_m,N)=1$, obtained by minimizing $B_\mathrm{min}$ over experimentally permissible values of $\tau_0$, $\tau_m$ and $N$ at fixed $N_s\mathcal{R}$ and $\tau_\mathrm{acq}$.
The spin-correlation time $\tau_m$ can be experimentally reduced by using scrambling pulses to re-randomize the spin ensemble \cite{Degen2007}. 
Accordingly, the optimization over $\tau_m$ is restricted to values below the maximum correlation time of 68~ms obtained in our measurements.
The details of the optimization are provided in the Supplemental Material \cite{Supplements}.
Note that unlike sensitivities defined from the local slope of the signal, this definition remains valid in the general nonlinear response regime of the sensor spin to the field.

\begin{figure}[h!]
    \centering
    \includegraphics[width=0.93\linewidth]{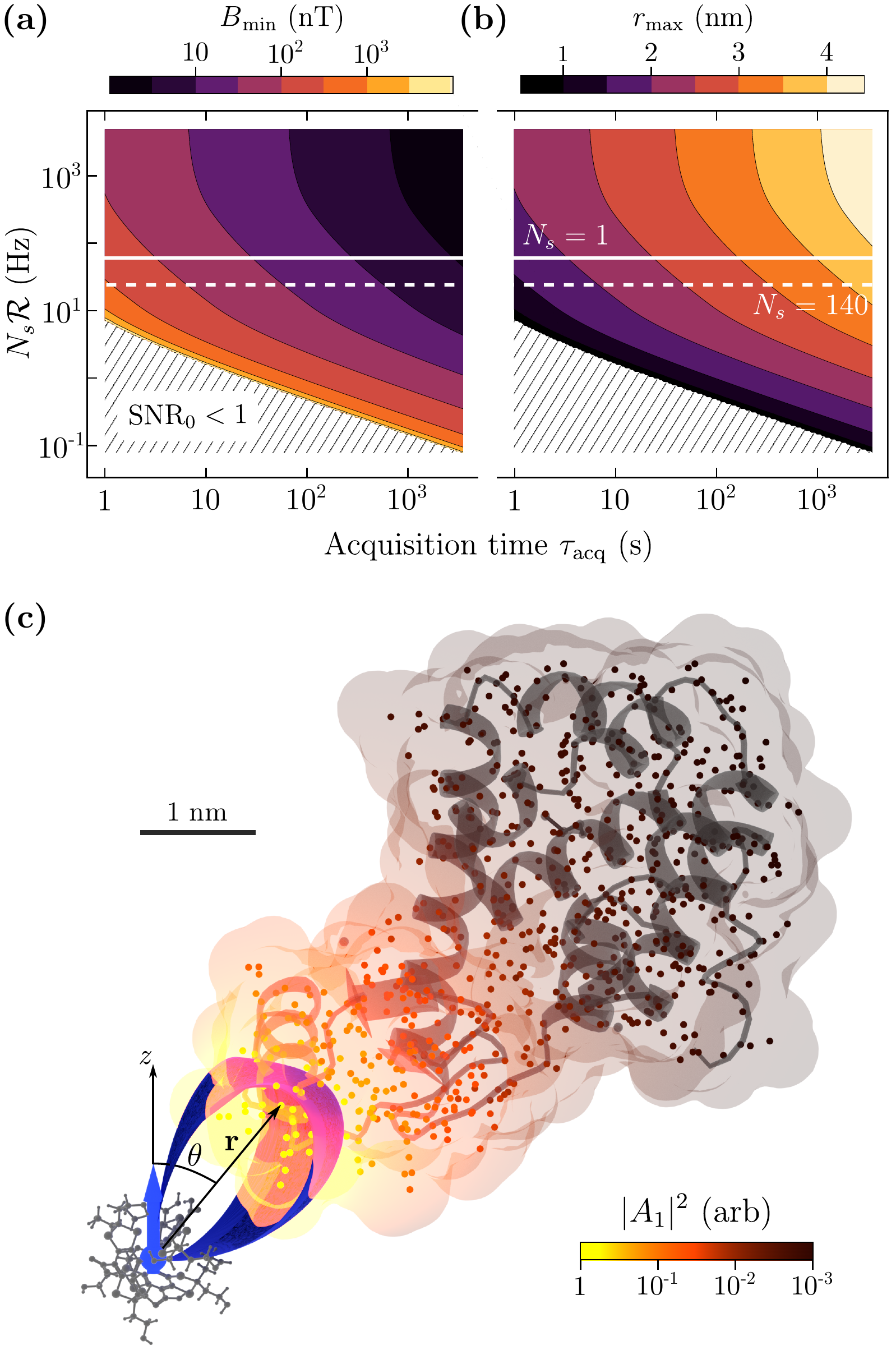}
    \caption{Magnetic field and nuclear spin sensitivity.
    (a) Calculated minimum detectable longitudinal field amplitude $B_{\min}$ as a function of the ensemble force-noise equivalent bandwidth 
    $N_s\mathcal{R}$ and acquisition time $\tau_{\mathrm{acq}}$, in order to achieve unit SNR.
    (b) Maximum sensor-nucleus separation $r_{\max}$ for unit SNR detection of a single proton spin using the $(t=0)$ XYXYd correlation spectroscopy sequence.
    The calculation assumes optimal orientation of the sensor-nucleus separation vector $\rr$ relative to the static field.
    The hatched region in (a,b) indicates parameter regimes for which the SNR of the bare electron signal satisfies $\mathrm{SNR}_0 < 1$, making sensing with unit SNR unfeasible.
    In both panels, the white dashed line at $N_s\mathcal{R} = 25$~Hz indicates the experimental readout parameters for the $\orderof 140$ sensor-spin ensemble in this work ($\mathcal{R} = 0.18$~Hz per spin), and the solid line indicates the projected parameters achievable with a single sensor spin using near-term improvements to the mechanical readout ($\mathcal{R} = 62$~Hz).
    (c) Envisioned SQUINT geometry for nuclear-spin sensing in a protein. The schematic shows a lysozyme target with a sensor radical positioned $\orderof 1~\mathrm{nm}$ from the protein; ${}^{13}$C nuclei are indicated by dots. The blue surface shows the $\abs{\sin(2\theta)}$ angular dependence of the pseudo-secular hyperfine coupling, where $\theta$ is the angle between the sensor--nucleus displacement vector $\rr$ and the static field. The color map within the protein indicates the relative contribution of each nuclear spin to the zero-delay correlation-spectroscopy signal, $\abs{A_1}^2 \propto \sin^2(2\theta)/r^6$.
    }
    \label{fig5}
\end{figure}

Figure~\ref{fig5}(a) shows a plot of the calculated $B_{\mathrm{min}}(N_s\mathcal{R},\tau_\mathrm{acq})$ detection threshold.
The calculation assumes the same XYXYd parameters used in Section~\ref{sec:extField}, i.e. $\Gamma = 0.34$ (index $k=1$), $T = 1.78~\mu$s and $T_d = 104~\mu$s.

For nuclear spin sensing, we consider a model in which each sensor electron is coupled to a single ${}^1$H spin at separation vector $\rr$.
The analysis is analogous to the external field case.
We consider the XYXYd-$N$ correlation spectroscopy sequence shown in Fig.~\ref{fig4}(e), evaluated at zero correlation delay ($t=0$), resulting in a signal of the form $\sin^2(NT \Gamma A_1/2)\, e^{-2NT/T_d}$ (Supplemental Material \cite{Supplements}).
The SNR of the measurement is
\begin{align}
\mathrm{SNR}(A_1,\tau_0,\tau_m,N)
&=
\sqrt{\frac{\tau_\mathrm{acq}}{2NT + \tau_0 + \tau_\mathrm{ovh}}}
 \times \nonumber\\[5pt]
 & \hspace{17pt} \frac{\sin^2(NT \Gamma A_1/2)\, e^{-2NT/T_d}}{\sigma(N_s\mathcal{R}, \tau_0,\tau_m)}.
\end{align}
The nuclear-spin detection threshold $A_{1,\min}(N_s\mathcal{R},\tau_\mathrm{acq})$ is defined as the smallest coupling strength for which $\mathrm{SNR}(A_{1,\min},\tau_0,\tau_m,N)=1$, obtained by minimizing $A_{1,\min}$ over experimentally permissible values of $\tau_0$, $\tau_m$ and $N$ at fixed $N_s\mathcal{R}$ and $\tau_\mathrm{acq}$.
Using a point-dipole approximation, we convert $A_{1,\min}$ to a maximum sensor-nucleus separation $r_\mathrm{max} = \left[3\mu_0 \hbar \gamma_H \gamma_e/(8\pi A_{1,\min})\right]^{1/3}$, where we assume optimal orientation of the hyperfine tensor, i.e. $\vert\sin(2\theta)\vert=1$.
Figure~\ref{fig5}(b) depicts the calculated $r_{\max}(N_s\mathcal{R},\tau_\mathrm{acq})$ for the same XYXYd parameters used in the external field case.

The results in Fig.~\ref{fig5}(a,b) illustrate the general dependence of the external-field and nuclear-spin detection thresholds on $N_s\mathcal{R}$ and the acquisition time $\tau_\mathrm{acq}$.
In both cases, larger $N_s\mathcal{R}$ improves the sensitivity and shifts the optimum toward shorter $\tau_0$ and $\tau_m$. 
Shorter $\tau_0$ reduces the duration of each measurement shot, while shorter $\tau_m$ allows more independent statistical-spin configurations to be sampled within the acquisition time.
At large $N_s\mathcal{R}$, further improvement is increasingly limited by the minimum accessible $\tau_0$ and the per-shot overhead (Supplemental Material \cite{Supplements}).
The hatched regions in Fig.~\ref{fig5}(a,b) indicate $(\tau_{\mathrm{acq}}, N_s\mathcal{R})$ pairs for which the SNR of the bare electron signal is less than unity ($\mathrm{SNR}_0 < 1$), making sensing with unit SNR unfeasible.

The dashed horizontal line in Fig.~\ref{fig5}(a,b) marks $N_s\mathcal{R} = 25$~Hz obtained from the experimental values $\mathcal{R} = 0.18$~Hz and $N_s = 140$ in this work. This corresponds to a mechanical force noise $S_F^{1/2}= 2.4~\mathrm{aN}/\sqrt{\mathrm{Hz}}$, $(\overline{G^2})^{1/2} = 1.6\times10^{5}~\mathrm{T/m}$ readout gradient, and $D = 0.98$.
With these parameters, a minimum detectable longitudinal field of $B_{\min} = 31$~nT and a maximum sensing distance of $r_{\max} = 2.5$~nm can be achieved with 1~min of averaging using the $140$-spin ensemble.

Although we did not optimize the readout parameters in the present measurements, the platform can support substantially larger $\mathcal{R}$ with straightforward improvements. 
Previous work demonstrated smaller-diameter SiNW arrays with force noise as low as $S_F^{1/2} = 500~\mathrm{zN}/\sqrt{\mathrm{Hz}}$, achieved through surface passivation and removal of the native oxide by vapor hydrofluoric acid, which reduced laser absorption and enabled operation closer to the 4~K bath temperature \cite{Sahafi2020}.
In addition, prior work has realized CFFGS devices that sustain 1.5$\times$ higher current density, which would increase the readout gradient to $\orderof 6\times 10^5~\mathrm{T/m}$ at 70~nm from the surface \cite{Haas2022}.

With these improvements, a $D = 1$ duty cycle would result in $\mathcal{R} = 62$~Hz [solid line in Fig.~\ref{fig5}(a,b)].
This would enable detection of a $B_{\min} = 22~\mathrm{nT}$ external field for $\tau_{\mathrm{acq}} = 1$~min of acquisition using a single sensor spin.
In addition, XYXYd correlation spectroscopy would allow detection of a single ${}^1$H spin at a maximum separation of $r_{\max} = 2.7~\mathrm{nm}$.

Figure~\ref{fig5}(c) illustrates an envisioned use of the SQUINT platform, in which a molecular spin sensor is positioned near a protein to probe local nuclear spins using XYXYd correlation spectroscopy. Spatial information could be obtained by Fourier encoding the target nuclei with large, time-dependent magnetic-field gradients applied during the correlation delay. When combined with nuclear-spin decoupling, this approach could enable angstrom-scale resolution, as demonstrated previously~\cite{Haas2022}. Multiple sensors could further be used to interrogate distinct regions of the protein, with individual sensors resolved through analogous Fourier encoding. 
Together, these capabilities would enable spatial mapping of nuclear spins in protein targets.

%%%%%%%%%%%%%%%%%%%%% Conclusion %%%%%%%%%%%%%%%%%%%%%%%%%%%%%%
\section{Conclusion}

We have presented SQUINT---a nanoscale quantum-sensing framework that combines molecular electron spins, ultra-sensitive mechanical spin readout, and high-fidelity dynamical decoupling.
Using dipolar decoupling with the XYXYd sequence, we achieve coherence times of $\orderof 400~\mu\mathrm{s}$ in a nanoscale ensemble of $\sim 140$ electron spins.
These capabilities enable frequency-selective sensing of nanotesla-scale RF magnetic fields, as well as detection and spectroscopy of local nuclear-spin ensembles, including ${}^1$H and $\sim 36$ naturally abundant ${}^{13}$C spins that dominate the observed signal.

In the present measurements, these nuclear signals originate from nuclei within the OX063 radical itself; however, synthesizing isotopically purified radicals using other isotopes (e.g., $^{12}$C, $^2$H) to suppress the probe's intrinsic nuclear-spin bath \cite{Jeschke2025} would reduce background contributions and extend the same protocol to nuclei in an external target molecule.
Notably, the readout parameters demonstrated here support operation with small sensor-spin ensembles, enabling unit-SNR detection of $\orderof 40$ electron spins in $1$~s of acquisition, corresponding to an OX063 concentration of $\orderof 60~\mu\mathrm{M}$ for a $1$~aL detection volume.
With the near-term improvements in mechanical readout quantified in Section~\ref{sec:sensitivity}, the platform is projected to reach single-electron sensitivity to nanotesla-scale magnetic fields and to detect individual ${}^1$H spins at nanoscale separations.

While the experiments here used SiNW force sensors and trityl-OX063 radicals, the SQUINT platform is not specific to either one.
State-of-the-art force transducers such as membrane and string resonators \cite{Prumbaum2025, Gisler2022, Reinhardt2016} can be integrated into the same sensing framework for enhanced force sensitivity.
Similarly, the platform is compatible with a broad range of molecular spin systems, including chemically engineered molecular qubits with extended coherence times obtained by tailoring the local nuclear spin environment \cite{Zadrozny2015,Atzori2016}.

A core feature of our approach is the combination of long-lived sensor coherence with large, time-dependent magnetic-field gradients built into the platform.
This enables direct integration of the present sensing framework with Fourier-based spatial encoding of either nuclear or electron spins, opening new possibilities for spatially resolved magnetic resonance spectroscopy and imaging on the molecular scale.

The use of molecular systems enables chemical control over sensor placement and environment, without being constrained by the geometry of a host crystal or proximity to a surface.
This allows sensor spins to be positioned in close and flexible arrangements relative to a target molecule, with sensing performance optimized through radical choice, functionalization, and engineering of the surrounding matrix.
Together, these capabilities establish a versatile and chemically tunable framework for nanoscale quantum sensing, enabling molecular-scale magnetic resonance measurements in complex environments beyond the constraints of defect-based sensors.

\begin{acknowledgments}
This work was undertaken thanks in part to funding from the Canada First Research Excellence
Fund (CFREF), and the Natural Sciences and Engineering Research Council of Canada (NSERC).
The University of Waterloo's QNFCF facility was used for this work. This infrastructure would not be possible without the significant contributions of CFREF-TQT, CFI, Industry Canada, the Ontario Ministry of Research and Innovation and Mike and Ophelia Lazaridis. Their support is gratefully acknowledged.
We would like to thank T. W. Borneman for helpful discussions and D. G. Cory for providing the trityl-OX063 samples, and useful comments.
\end{acknowledgments}

% \appendix

% \nocite{*}

% \bibliography{ref}% Produces the bibliography via BibTeX.

\providecommand{\noopsort}[1]{}\providecommand{\singleletter}[1]{#1}%
%

%%% Supplements %%%
\clearpage
\clearpage


\section*{Supplementary material}

# Supplemental Material for Long-Lived Mechanically-Detected Molecular Spins for Quantum Sensing

Sahand Tabatabaei,[1,2] Pritam Priyadarsi,[1,2] Daniel Tay,[1,2] Namanish Singh,[1,2] Pardis Sahafi,[1,2] Andrew Jordan,[1,2] and Raffi Budakian[1,2,*]

[1]*Department of Physics and Astronomy, University of Waterloo, Waterloo, ON, Canada, N2L3G1*
[2]*Institute for Quantum Computing, University of Waterloo, Waterloo, ON, Canada, N2L3G1*
(Dated: March 4, 2026)

## I. SAMPLE PREPARATION AND ATTACHMENT

This section summarizes the preparation and attachment of the polystyrene spacer and the OX063-containing sugar matrix used in the experiments. All steps were carried out using the same microinjector-micromanipulator setup described in the Supplemental Material for Ref. [1].

### A. Polystyrene tip spacer

To decouple the OX063 spins from fast-relaxing paramagnetic defects on the silicon nanowire (SiNW) [1], we introduced a polystyrene spacer. The spacer was made from a solution prepared using the constituents listed in Table I. The solution was loaded into a glass capillary, and left to dissolve for 5 days, with occasional stirring. The setup was then left undisturbed for 50 minutes at 25 °C to allow partial toluene evaporation, increasing the polystyrene concentration at the droplet surface and improving adhesion to the SiNW tip. A droplet at the capillary tip was brought into contact with the SiNW tip under an optical microscope, and then slowly withdrawn to pull a polystyrene spacer of length $\sim 3$ μm.

TABLE I. Polystyrene spacer solution constituents.

| Constituent | Amount |
| --- | --- |
| Diethyl phthalate; $C_{12}H_{14}O_4$ | 15 mL |
| Toluene; $C_6H_5CH_3$ | 15 mL |
| Polystyrene; $(C_8H_8)_n$ (molecular weight 35,000) | 8.3 g |

### B. Sample solution preparation

An OX063 stock solution was prepared using the constituents listed in Table II. The sugars and OX063 radicals were combined in a vial, $D_2O$ was added, and the vial was placed on a vortex shaker to dissolve the contents. The stock solution was formulated to achieve a 10:1 (w/w) ratio of trehalose dihydrate to sucrose, which ensures that the sugar mixture forms a glassy matrix at 4 K [2].

TABLE II. OX063 stock solution constituents.

| Constituent | Amount |
| --- | --- |
| Sucrose; $C_{12}H_{22}O_{11}$ | 518 mg |
| Trehalose dihydrate; $C_{12}H_{22}O_{11} \cdot 2H_2O$ | 5.22 g |
| $D_2O$ | 9.7 mL |
| Trityl-OX063 | 2.3 mg |

---

[*] rbudakian@uwaterloo.ca

## C. Sample attachment

After the polystyrene spacer was formed, the SiNW was stored in a dry nitrogen box for 2 days before sample attachment. Attachment was carried out in a controlled environment at 45% relative humidity and 25 °C. The sugar sample was drawn into a ground-glass pipette and dispensed using a pneumatic injector to form a droplet at the pipette tip. Under an optical microscope, the droplet was brought into contact with the polystyrene spacer tip and the pipette was withdrawn, leaving an OX063-containing sugar droplet at the end of the spacer.

## II. CFFGS FIELD DISTRIBUTION

The current-focusing field gradient source (CFFGS) is a $\sim 150$ nm-wide metallic constriction that generates the magnetic fields and field gradients used for spin control and detection. A detailed description of the fabrication process for the CFFGS is given in Section 2.1 of the Supplemental Material of Ref. [1]. To characterize the spatial distribution of the CFFGS fields, we used COMSOL finite element simulation software. The simulated electron Rabi frequency $u(\mathbf{r}) = \gamma_e \sqrt{B_x^2(\mathbf{r}) + B_y^2(\mathbf{r})}/(4\pi)$ and readout gradient $G(\mathbf{r}) = \partial B_z(\mathbf{r})/\partial y$ are shown in Fig. S1. We note that in our geometry, the contribution of the orthogonal gradient component $\partial B_z(\mathbf{r})/\partial x$ to the measured signal is negligible because (1) it is substantially smaller than $G(\mathbf{r})$ over the detection volume and (2) the associated force is approximately perpendicular to the principal axis of the SiNW mode used for spin detection.

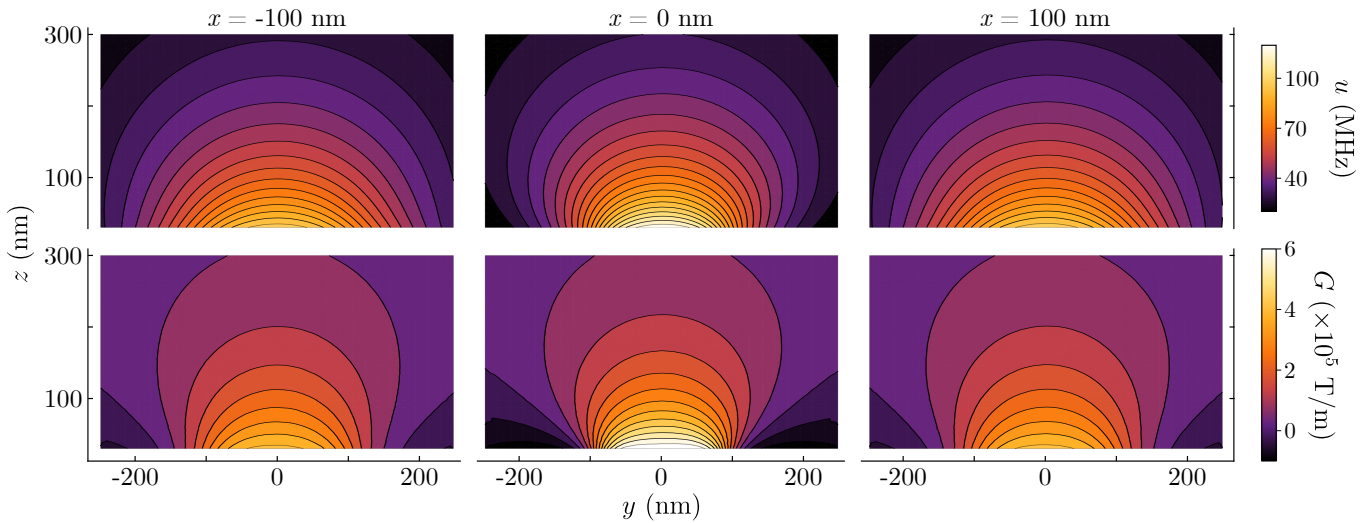

FIG. S1. CFFGS field distribution. Simulated $y$-$z$ cross sections of (top) the electron Rabi frequency $u(\mathbf{r})$ and (bottom) readout gradient $G(\mathbf{r}) = \partial B_z(\mathbf{r})/\partial y$. The $u(\mathbf{r})$ profiles were generated for a peak current of 5 mA through the CFFGS, while the $G(\mathbf{r})$ profiles were generated for a peak current of 45 mA. Coordinates are defined with the CFFGS current along $x$ and its top surface at $z = 0$, centered at $x = y = 0$.

## III. CFFGS DRIVE ELECTRONICS

Figure S2 summarizes the electronics used to generate and deliver the waveforms applied to the CFFGS. Signal generation is divided into three frequency bands: (1) an X-band microwave chain used for ESR control, (2) a MHz-range network for generating the AC drive currents used in the external-field sensing measurements, and (3) a sub-MHz network used to generate the gradient modulation at the SiNW frequency for MAGGIC readout. The external-field and gradient modulation waveforms are generated by the arbitrary waveform generators AWG1 and AWG2, respectively, and are selected using a TTL-controlled switch that internally terminates the unselected port with 50 Ω. These low-frequency waveforms are delivered differentially to the CFFGS, with an optional $V_{\rm DC}$ bias that can mitigate the effect of charges on the SiNW. The MHz-range chain is also compatible with generating NMR control pulses when needed (up to $\sim 100$ MHz).

Microwave (ESR) pulses are generated using an IQ single-sideband up-conversion scheme: a baseband I/Q waveform is digitally mixed to an intermediate frequency (IF) of 377 MHz and generated by AWG1. The IF signal is mixed

with an 8.883 GHz local oscillator (LO) and bandpass filtered (9.0–9.4 GHz) to suppress LO leakage and higher-order mixer products before amplification. To mitigate unwanted $T_1$ relaxation of the electron spins due to microwave amplifier noise, a fast high-power switch gates the microwave output when no pulses are being applied. Finally, the microwave and low-frequency waveforms are combined using diplexers and delivered to the CFFGS through 50 Ω coaxial lines inside the probe. The timing between the AWGs, TTL switching pulses, and the lock-in reference used for demodulating the interferometer signal is controlled by an FPGA (not shown).

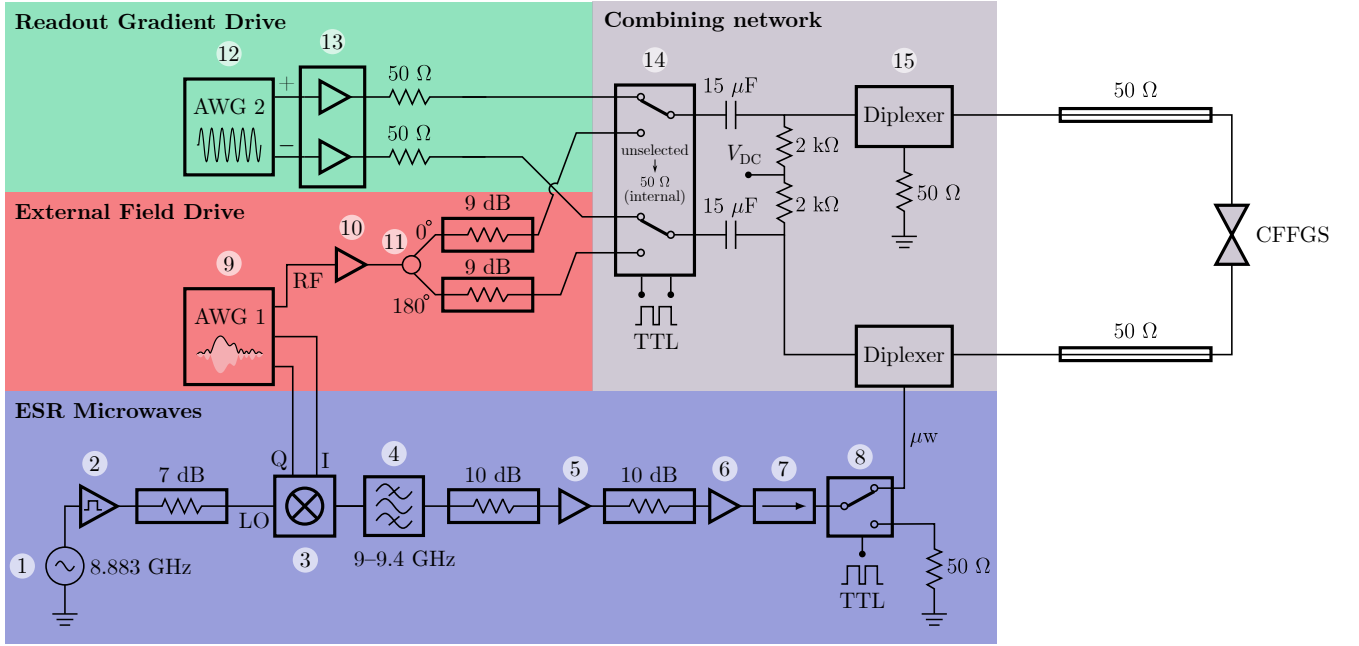

FIG. S2. CFFGS drive electronics.

## IV. ELECTRON SPIN COUNT ESTIMATION

To estimate the number of contributing electron spins, we continuously ran the MAGGIC protocol and recorded the SiNW displacement demodulated at the mechanical resonance $f_c = 297.069$ kHz using a lock-in amplifier. The lock-in was phase referenced to the MAGGIC waveform, such that the spin force contribution was entirely in the in-phase channel. Therefore, the in-phase channel contained both the statistical spin force contribution and the thermal force noise of the oscillator, while the quadrature channel contained only the oscillator noise contribution. Computing the (single-sided) PSDs $S_I(f)$ and $S_Q(f)$ of each channel (Fig. S3), we compared the spin force and thermal noise contributions to estimate the number of spins in the ensemble.

The PSDs are the sum of independent thermal and spin force contributions $S_I(f) = S_{\text{th}}(f) + S_{\text{spin}}(f)$ and $S_Q(f) = S_{\text{th}}(f)$. For all spin measurements, the SiNW was feedback-damped using a piezoelectric transducer to a quality factor $Q \simeq 1800$, corresponding to a ringdown time $\tau_Q = Q/(\pi f_c) \simeq 2$ ms [3]. The thermal force noise of the oscillator

produces identical Lorentzian spectra in both quadratures

$$S_{\text{th}}(f) = \alpha \frac{S_F}{1 + (2\pi\tau_Q f)^2}, \tag{S1}$$

where $S_F = 4k_B T_{\text{NW}} k/(2\pi f_c Q_0)$ is its thermal force noise PSD. Given the temperature $T_{\text{NW}} = 8$ K, spring constant $k = 0.7$ N/m, and intrinsic quality factor $Q_0 = 28{,}000$, we have $S_F^{1/2} = 2.4$ aN/$\sqrt{\text{Hz}}$. Here, the SiNW temperature $T_{\text{NW}}$ and spring constant $k$ were measured using the methods described in [4]. The proportionality constant $\alpha$ accounts for an overall force-to-voltage calibration that is common to both lock-in channels. Fluctuations of the spin force generate an additional in-phase contribution

$$S_{\text{spin}}(f) = \alpha\, N_s \mu^2 D^2 \overline{G^2} \frac{\tau_m}{1 + (2\pi\tau_m f)^2}, \tag{S2}$$

where $\mu = 9.27 \times 10^{-24}$ J/T is the electron magnetic moment, $N_s$ is the electron spin count, $\tau_m$ is the spin correlation time under MAGGIC, $D = 0.98$ is the MAGGIC duty cycle, and $\overline{G^2}$ is the sample-average of the squared readout gradient.

To extract the spin count, we first fit the measured $S_Q(f)$ to Eq. (S1) with fit parameters $A_{\text{th}} \equiv \alpha S_F$ and $\tau_Q$. We then isolate the spin contribution to the in-phase PSD by subtracting the fitted quadrature-channel PSD from $S_I(f)$. The result is subsequently fitted to Eq. (S2) with fit parameters $A_{\text{spin}} = \alpha \mu^2 D^2 \overline{G^2} N_s$ and $\tau_m$. Taking the ratio of the fitted amplitudes yields $N_s \overline{G^2} = A_{\text{spin}} S_F/(A_{\text{th}} \mu^2 D^2) = 3.5 \times 10^{12}$ (T/m)$^2$.

To estimate $\overline{G^2}$, we use the measured distributions of Rabi frequencies and longitudinal fields from the CFFGS [Fig. 1(e) and Fig. 3(e) of the main text, respectively] to constrain spatial distribution of OX063 radicals. Assuming azimuthal symmetry and an approximately uniform OX063 density, we represent the boundary of the distribution by a piecewise-linear radial profile whose vertices are adjusted for consistency with both measured distributions [Fig. S3(c,d)]. Since the OX063 radicals are deposited on the polystyrene shell, this reconstruction should be interpreted as the OX063 spatial distribution rather than the physical geometry of the droplet itself. Using finite-element simulations of the CFFGS field, we then compute $\overline{G^2} = \left(1.6 \times 10^5 \text{ T/m}\right)^2$ for a peak readout current of 45 mA by averaging $G^2(\mathbf{r}) = [\partial B_z(\mathbf{r})/dy]^2$ over the resulting OX063 distribution. Combining this with the $N_s \overline{G^2}$ value from the PSD fits results in an estimated electron spin count $N_s = 140$.

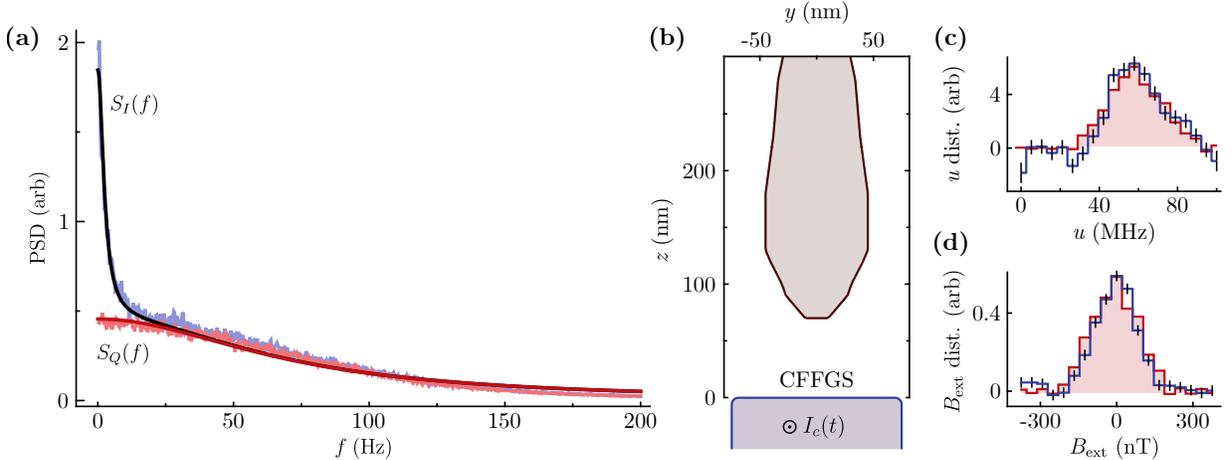

FIG. S3. Electron spin count estimation. (a) Single-sided power spectral density (PSD) of the demodulated SiNW displacement under continuous MAGGIC. Blue and red show the in-phase $S_I(f)$ and quadrature $S_Q(f)$ spectra. Thin lines are the measured data and thick solid curves are fits, yielding $\tau_Q = 2.25(2)$ ms and $\tau_m = 68(4)$ ms. The quadrature PSD $S_Q(f)$ is set solely by the oscillator's thermal force noise, while the in-phase PSD $S_I(f)$ includes both thermal and spin force contributions. (b) Cross section of reconstructed OX063 distribution constrained by the measured Rabi-frequency distribution (c) and longitudinal field distribution (d), with the bottom indicating the cross section of the CFFGS. Red traces in (c,d) are simulations based on the reconstructed OX063 distribution, and blue traces are the corresponding measured distributions from Fig. 1(e) and Fig. 3(e) of the main text.

## V. EFFECTIVE SENSING HAMILTONIAN

In this section, we derive the effective Hamiltonians under the XYXYd sequence used in Sections IV and V of the main text. We consider a single electron spin in both cases. Generalization to a non-interacting ensemble is straightforward.

### A. External Field Sensing

Let $\mathbf{B}(t)$ be the time-dependent external field applied during the XYXYd sequence. We assume that $\mathbf{B}(t)$ varies much slower than the electron Rabi frequency $u$, and that $u \gg \gamma_e |\mathbf{B}(t)|/(2\pi)$, to allow for average Hamiltonian theory [5]. The Hamiltonian in the electron rotating frame is $H(t) = -\gamma_e B_z(t) S^z + H_C(t)$, where $H_C(t) = 2\pi u\, \mathbf{a}(t) \cdot \mathbf{S}$ is the control Hamiltonian, and the transverse $B_x, B_y$ terms are omitted by the rotating-wave approximation. Transforming into the interaction frame of $H_C(t)$ (the toggling frame), we have

$$\tilde{H}(t) = -\gamma_e B_z(t)\, U_C^\dagger(t) S^z U_C(t), \tag{S3}$$

where $U_C(t) = \mathrm{Texp}[-i\int_0^t dt'\, H_C(t')]$ is the control unitary, with Texp denoting the time-ordered exponential. We write $U_C^\dagger(t) S^z U_C(t) = \mathbf{\Gamma}(t) \cdot \mathbf{S}$, which is $T$-periodic under repeated applications of an XYXYd sequence with duration $T$. Writing $\mathbf{\Gamma}$ as a Fourier series $\mathbf{\Gamma}(t) = \sum_{n=-\infty}^{\infty} \mathbf{c}_n\, e^{i2\pi nt/T}$ and calculating the average Hamiltonian to zeroth order gives

$$H_{\mathrm{eff}} = \frac{1}{NT} \int_0^{NT} dt\, \tilde{H}(t) = -\gamma_e \sum_{n=-\infty}^{\infty} \left( \int_{-\infty}^{\infty} dt\, w_n(t) B_z(t) \right) \mathbf{c}_n \cdot \mathbf{S}, \tag{S4}$$

where $w_n(t) = e^{i2\pi nt/T}/(NT)$ for $t \in [0, NT]$ and zero otherwise. Rewriting the integral in the frequency domain using the Plancherel identity results in

$$\int_{-\infty}^{\infty} dt\, w_n(t) B_z(t) = (-1)^{nN} \int_{-\infty}^{\infty} d\nu\, \mathrm{sinc}\left[NT\left(\nu - \frac{n}{T}\right)\right] e^{i\pi NT\nu} \tilde{B}_z(\nu), \tag{S5}$$

and thus

$$H_{\mathrm{eff}} = -\gamma_e \sum_{n=-\infty}^{\infty} (-1)^{nN} \left( \int_{-\infty}^{\infty} d\nu\, \mathrm{sinc}\left[NT\left(\nu - \frac{n}{T}\right)\right] e^{i\pi NT\nu} \tilde{B}_z(\nu) \right) \mathbf{c}_n \cdot \mathbf{S}, \tag{S6}$$

Eq. S6 can be simplified using symmetries of the $\mathbf{\Gamma}(t)$ trajectory:

1. $\Gamma_x(t)$ and $\Gamma_y(t)$ have spectral content only at frequencies $\sim u$, so $c_n^x \simeq c_n^y \simeq 0$ over the support of $\tilde{B}_z(\nu)$.

2. $\Gamma_z(t)$ has quarter-wave symmetry, i.e. $\Gamma_z(t + nT/4) = (-1)^n \Gamma_z(t)$, meaning that the only non-zero Fourier coefficients are at frequencies $\nu_k = 2(2k-1)/T$, where $k \in \mathbb{Z}$.

An example of the $\mathbf{\Gamma}(t)$ trajectory is depicted in Fig. S4. Combining these properties, we get the final effective Hamiltonian $H_{\mathrm{eff}} = -\Omega_N(T) S^z$, with

$$\Omega_N(T) = \gamma_e \int_{-\infty}^{\infty} d\nu\, \zeta_N(T\nu) \tilde{B}_z(\nu) e^{i\pi NT\nu}, \tag{S7}$$

and

$$\zeta_N(T\nu) = \sum_{k=-\infty}^{\infty} \Gamma_k\, \mathrm{sinc}\left[NT(\nu - \nu_k)\right], \tag{S8}$$

where $\Gamma_k \equiv c_{2(2k-1)}^z$. Note that since $U_C(T) = \mathbb{1}$, the expression for $H_{\mathrm{eff}}$ also describes the rotating frame evolution at the end of the sequence.

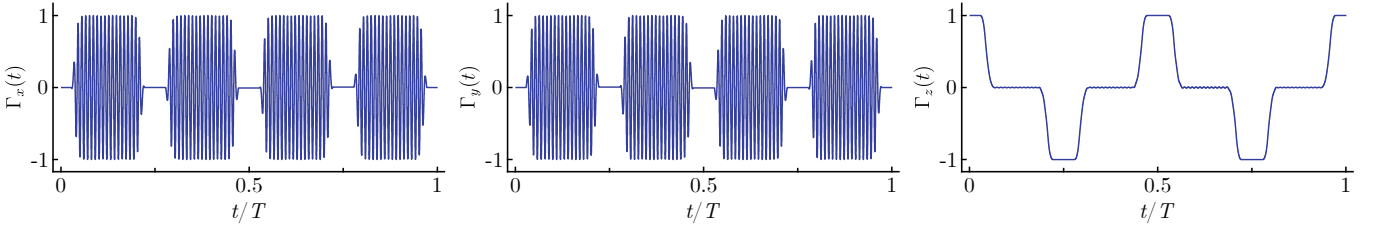

FIG. S4. Calculated $\mathbf{\Gamma}(t)$ trajectory for an XYXYd sequence with $T = 1.78$ μs, for an electron at Rabi frequency $u = 55$ MHz.

## B. Nuclear Spin Sensing

Derivation of the effective Hamiltonian for nuclear spin sensing is closely analogous to the external field case. We assume spin-$I$ nuclei with Larmor frequency $\nu_\text{nuc}$. In the electron rotating frame, the Hamiltonian is

$$H(t) = -2\pi \sum_l \nu_\text{nuc} I_l^z + \sum_l \left(A_{0l} I_l^z + \mathbf{A_{1l}} \cdot \mathbf{I}_l\right) S^z + H_C(t). \tag{S9}$$

We assume that $2\pi u \gg 2\pi\nu_\text{nuc}, |A_{0l}|, |A_{1l}|$ to allow for average Hamiltonian theory, and write the pseudo-secular coupling vector as $\mathbf{A}_{1l} = A_{1l} \left(\cos\alpha_l \, \hat{\mathbf{x}} + \sin\alpha_l \, \hat{\mathbf{y}}\right)$. Transforming into the interaction frame of the $-2\pi \sum_l \nu_\text{nuc} I_l^z + H_C(t)$ term gives

$$\tilde{H}(t) = \sum_l \left[ A_{0l} I_l^z + \frac{A_{1l}}{2} \left( e^{-i(2\pi\nu_\text{nuc} t + \alpha_l)} I_l^+ + e^{i(2\pi\nu_\text{nuc} t + \alpha_l)} I_l^- \right) \right] \mathbf{\Gamma}(t) \cdot \mathbf{S}, \tag{S10}$$

where $\mathbf{\Gamma}(t) \cdot \mathbf{S} = U_C^\dagger(t) S^z U_C(t)$ with $\mathbf{\Gamma}(t)$ being $T$-periodic, similar to the external field case. Writing $\mathbf{\Gamma}(t)$ as a Fourier series, accounting for the symmetries outlined in Section V A, and averaging over $[0, NT]$ gives

$$H_\text{eff} = \sum_l A_{1l} \sum_{k=-\infty}^{\infty} \text{sinc}\left[NT\left(\nu_\text{nuc} - \nu_k\right)\right] \frac{1}{2} \left( e^{-i(\pi NT\nu_\text{nuc} + \alpha_l)} I_l^+ + e^{i(\pi NT\nu_\text{nuc} + \alpha_l)} I_l^- \right) \Gamma_k S^z. \tag{S11}$$

The $A_{0l}$ term averages to zero since $\mathbf{\Gamma}(t)$ lacks a DC component. This gives the final effective Hamiltonian

$$H_\text{eff} = \sum_l \zeta_N(T\nu_\text{nuc}) A_{1l} \left[\hat{\boldsymbol{\phi}}(\pi NT\nu_\text{nuc} + \alpha_l) \cdot \mathbf{I}_l\right] S^z, \tag{S12}$$

where $\zeta_N(T\nu)$ is defined in Eq. S8 and we introduce the shorthand $\hat{\boldsymbol{\phi}}(\varphi) \equiv \cos\varphi \, \hat{\mathbf{x}} + \sin\varphi \, \hat{\mathbf{y}}$. Eq. S12 is equivalent to effective Hamiltonian $H_\text{eff} = \sum_l \zeta_N(T\nu_\text{nuc}) A_{1l} I_l^x S^z$ used in the main text, up to a re-definition of the transverse nuclear axes. Transforming back into the electron rotating frame, evolution under XYXYd-$N$ is described by the propagator

$$U_\text{eff}(N, T) = e^{i2\pi NT \sum_l \nu_\text{nuc} I_l^z} e^{-iNT H_\text{eff}}. \tag{S13}$$

## VI. ELECTRON SIGNAL UNDER THE SENSING HAMILTONIAN

The effective Hamiltonians derived in Section V can be used to derive analytic expressions for the measured electron signal under the sensing measurements. Throughout this section, we normalize the signal $C$ such that $C = 1$ in the absence of any control sequence (i.e., for the bare electron signal).

### A. External Field Sensing

#### 1. Direct XYXYd Measurements

For direct XYXYd measurements [Fig. 3(a–e) of the main text], the control sequence is $\text{AHP}_{\hat{\mathbf{z}} \to \hat{\mathbf{x}}}$ – XYXYd-$N$ – $\text{AHP}_{\hat{\mathbf{n}} \to \hat{\mathbf{z}}}$, where $\hat{\mathbf{n}} = \hat{\mathbf{x}}$ ($\hat{\mathbf{y}}$) for the in-phase (quadrature) measurement. Using the effective Hamiltonian $H_\text{eff} = $

$-\Omega_N(T)S^z$ [Eq. S7], the corresponding propagators for a single electron are $U_I = U_{\hat{\mathbf{x}} \to \hat{\mathbf{z}}} e^{-iNT\Omega_N(T)S^z} U_{\hat{\mathbf{z}} \to \hat{\mathbf{x}}}$ and $U_Q = U_{\hat{\mathbf{y}} \to \hat{\mathbf{z}}} e^{-iNT\Omega_N(T)S^z} U_{\hat{\mathbf{z}} \to \hat{\mathbf{x}}}$. Thus, the in-phase signal is

$$C_I \propto \mathrm{Tr}\left[S^z U_I^\dagger S^z U_I\right] = \mathrm{Tr}\left[S^x e^{-iNT\Omega_N(T)S^z} S^x e^{iNT\Omega_N(T)S^z}\right], \tag{S14}$$

Evaluating Eq. S14 and similarly calculating the quadrature signal gives

$$C_I = \cos\left[NT\Omega_N(T)\right], \qquad C_Q = \sin\left[NT\Omega_N(T)\right]. \tag{S15}$$

### 2. Correlation Spectroscopy Measurements

For the correlation spectroscopy sequence of Fig. 3(g) of the main text, the control sequence is

$$\mathrm{AHP}_{\hat{\mathbf{z}} \to \hat{\mathbf{x}}} - \mathrm{XYXYd}\text{-}N - \mathrm{AHP}_{\hat{\mathbf{y}} \to \hat{\mathbf{z}}} - \text{correlation delay } (t) - \mathrm{AHP}_{\hat{\mathbf{z}} \to \hat{\mathbf{x}}} - \mathrm{XYXYd}\text{-}N - \mathrm{AHP}_{\hat{\mathbf{y}} \to \hat{\mathbf{z}}},$$

where the correlation delay stores the phase of the first block in the longitudinal spin component. Given the $T_2^* = 78$ ns electron dephasing time, we model electron dephasing during the correlation delay with a completely dephasing map $\mathcal{E}$, acting as $\mathcal{E}(O) \equiv \Pi_+ O \Pi_+ + \Pi_- O \Pi_-$, where $\Pi_\pm$ are the eigenprojectors of $S^z$. The correlation signal is

$$C(t) \propto \mathrm{Tr}\left[S^z\, U_N^\dagger(0)\, \mathcal{E}\left(U_N^\dagger(t+NT) S^z U_N(t+NT)\right) U_N(0)\right], \tag{S16}$$

where $U_N(\tau) \equiv U_{\hat{\mathbf{y}} \to \hat{\mathbf{z}}}\, e^{iNT\Omega_N(\tau,T)S^z} U_{\hat{\mathbf{z}} \to \hat{\mathbf{x}}}$, and

$$\Omega_N(\tau,T) = \gamma_e \int_{-\infty}^{\infty} d\nu\, \zeta_N(T\nu) \tilde{B}_z(\nu) e^{i2\pi\nu\tau} e^{i\pi NT\nu}. \tag{S17}$$

The phase factor $e^{i2\pi\nu\tau}$ is introduced to account for a time shift $\tau$ between the start of the second XYXYd block and the field modulation. Using the action of the AHPs to rotate $S^z$ into transverse components, and using the fact that $\mathcal{E}$ removes all terms proportional to $S^x$ and $S^y$, Eq. S16 reduces to

$$C(t) = \sin\left[NT\Omega_N(0,T)\right]\sin\left[NT\Omega_N(t+NT,T)\right]. \tag{S18}$$

In the small-phase regime $|NT\Omega_N(\tau,T)| \ll 1$, Eq. S18 becomes $C(t) = N^2T^2\Omega_N(0,T)\Omega_N(t+NT,T)$. The signal dependence on $t$ is entirely in $\Omega_N(t+NT,T) = \gamma_e \int d\nu\, \zeta_N(T\nu)\tilde{B}_z(\nu)e^{i2\pi\nu(t+3NT/2)}$. Thus, measuring $C(t)$ and Fourier transforming yields a spectrum of the external field within the passband of the XYXYd filter function.

### B. Nuclear Spin Sensing

#### 1. Direct XYXYd Measurements

Similar to the external field case, the signal under direct XYXYd measurements is

$$C \propto \mathrm{Tr}\left[S^z U_{\hat{\mathbf{z}} \to \hat{\mathbf{x}}}^\dagger U_{\mathrm{eff}}^\dagger(N,T) U_{\hat{\mathbf{x}} \to \hat{\mathbf{z}}}^\dagger S^z U_{\hat{\mathbf{x}} \to \hat{\mathbf{z}}} U_{\mathrm{eff}}(N,T) U_{\hat{\mathbf{z}} \to \hat{\mathbf{x}}}\right] = \mathrm{Tr}\left[S^x U_{\mathrm{eff}}^\dagger(N,T) S^x U_{\mathrm{eff}}(N,T)\right], \tag{S19}$$

where $U_{\mathrm{eff}}(N,T)$ is given in Eq. S13. Defining $\theta_l \equiv NT\zeta_N(T\nu_{\mathrm{nuc}})A_{1l}$, the effective propagator $U_{\mathrm{eff}}(N,T)$ rotates the transverse electron operator about $S^z$ by the nuclear phase $\Phi_0 \equiv \sum_l \theta_l \hat{\phi}(\pi NT\nu_{\mathrm{nuc}} + \alpha_l) \cdot \mathbf{I}_l$. Using Eq. S19 and the cyclic property of the trace, only the longitudinal component of this rotation contributes, giving

$$C \propto \mathrm{Tr}\left[\cos\Phi_0\right] = \mathrm{Re}\left(\mathrm{Tr}[e^{i\Phi_0}]\right) = \mathrm{Re}\left(\prod_l \mathrm{Tr}\left[e^{i\theta_l \hat{\phi}(\pi NT\nu_{\mathrm{nuc}}+\alpha_l)\cdot \mathbf{I}_l}\right]\right), \tag{S20}$$

Realizing that $\mathrm{Tr}\left[e^{i\theta_l \hat{\phi}(\pi NT\nu_{\mathrm{nuc}}+\alpha_l)\cdot \mathbf{I}_l}\right] = \sum_{m=-I}^{I} e^{i\theta_l m} = \sin\left(\frac{2I+1}{2}\theta_l\right)/\sin\left(\frac{1}{2}\theta_l\right)$ and substituting $\theta_l = NT\zeta_N(T\nu_{\mathrm{nuc}})A_{1l}$ leads to the final expression

$$C = \frac{1}{(2I+1)^{N_{\mathrm{nuc}}}}\prod_l \frac{\sin\left(\frac{2I+1}{2}NT\zeta_N(T\nu_{\mathrm{nuc}})A_{1l}\right)}{\sin\left(\frac{1}{2}NT\zeta_N(T\nu_{\mathrm{nuc}})A_{1l}\right)}, \tag{S21}$$

where $N_{\mathrm{nuc}}$ is the number of nuclei coupled to the electron. Evaluating Eq. S21 for $I = 1/2$ leads to Eq. 15 of the main text.

### 2. Correlation Spectroscopy Measurements

For correlation spectroscopy, we have

$$C(t) \propto \mathrm{Tr}\left[S^z\, U_e^\dagger(N,T) U_0^\dagger(t) \mathcal{E}\left(U_e^\dagger(N,T) S^z U_e(N,T)\right) U_0(t) U_e(N,T)\right], \tag{S22}$$

where $U_e(N,T) \equiv U_{\hat{\mathbf{y}} \to \hat{\mathbf{z}}} U_{\mathrm{eff}}(N,T) U_{\hat{\mathbf{z}} \to \hat{\mathbf{x}}}$ and $U_0(t) \equiv e^{-it\sum_l(-2\pi\nu_{\mathrm{nuc}} + A_{0l}S^z)I_l^z}$ describes the free evolution time. Using the action of the AHP rotations and standard spin-rotation identities, the signal reduces to

$$C(t) \propto \mathrm{Tr}\left[\sin\Phi_1 \sin\Phi_2\right], \tag{S23}$$

where $\Phi_1 \equiv \sum_l \theta_l \hat{\phi}(\alpha_l - \pi NT\nu_{\mathrm{nuc}}) \cdot \mathbf{I}_l$ and $\Phi_2 \equiv \sum_l \theta_l\, \hat{\phi}[\alpha_l + (\pi NT + 2\pi t)\nu_{\mathrm{nuc}} - tA_{0l}S^z] \cdot \mathbf{I}_l$ are the phase operators in the first and second XYXYd-$N$ block, respectively. Thus, $C(t)$ is the correlation of the two accumulated phases. Evaluating the trace over the electron $m = \pm 1/2$ subspace and writing the sine functions with complex exponentials gives

$$C(t) \propto \frac{1}{2}\sum_{m=\pm 1/2} \mathrm{Re}\left(\mathrm{Tr}\left[e^{i\Phi_1}e^{-i\Phi_{2m}}\right] - \mathrm{Tr}\left[e^{i\Phi_1}e^{i\Phi_{2m}}\right]\right), \tag{S24}$$

where $\Phi_{2m}$ is obtained from $\Phi_2$ by replacing $S^z \to m$. Since $\Phi_1$ and $\Phi_{2m}$ are sums of commuting single-nucleus operators, the nuclear trace factorizes as $\mathrm{Tr}\left[e^{i\Phi_1}e^{\pm i\Phi_{2m}}\right] = \prod_l T_{lm}^{(\pm)}$, where

$$T_{lm}^{(\pm)} \equiv \mathrm{Tr}\left[e^{i\theta_l\hat{\phi}(\alpha_l - \pi NT\nu_{\mathrm{nuc}})\cdot\mathbf{I}_l} e^{\pm i\theta_l\hat{\phi}[\alpha_l + (\pi NT + 2\pi t)\nu_{\mathrm{nuc}} - tA_{0l}m]\cdot\mathbf{I}_l}\right]. \tag{S25}$$

The product of the two rotations is a single rotation with angle $\gamma_{lm}^{(\pm)}$, where

$$\cos\left(\frac{\gamma_{lm}^{(\pm)}}{2}\right) = \cos^2\left(\frac{\theta_l}{2}\right) \mp \cos\left[2\pi\nu_{\mathrm{nuc}}(t+NT) - tA_{0l}m\right]\sin^2\left(\frac{\theta_l}{2}\right). \tag{S26}$$

Hence $T_{lm}^{(\pm)} = \sum_{r=-I}^{I} e^{ir\gamma_{lm}^{(\pm)}} = \sin\left(\frac{2I+1}{2}\gamma_{lm}^{(\pm)}\right) / \sin\left(\frac{1}{2}\gamma_{lm}^{(\pm)}\right)$. Substituting $\mathrm{Tr}\left[e^{i\Phi_1}e^{\pm i\Phi_{2m}}\right] = \prod_l T_{lm}^{(\pm)}$ into Eq. S24 yields

$$C(t) = \frac{1}{4(2I+1)^{N_{\mathrm{nuc}}}} \sum_{m=\pm 1/2}\left[\prod_l \frac{\sin\left(\frac{2I+1}{2}\gamma_{lm}^{(-)}\right)}{\sin\left(\frac{1}{2}\gamma_{lm}^{(-)}\right)} - \prod_l \frac{\sin\left(\frac{2I+1}{2}\gamma_{lm}^{(+)}\right)}{\sin\left(\frac{1}{2}\gamma_{lm}^{(+)}\right)}\right], \tag{S27}$$

where $\gamma_{lm}^\pm$ is defined implicitly by Eq. S26. Equation S27 can be simplified in the small-phase regime to leading order in $\theta_l = NT\zeta_N(T\nu_{\mathrm{nuc}})A_{1l}$:

$$C(t) \simeq \frac{I(I+1)}{3} N^2 T^2 \sum_l \zeta_N^2(T\nu_{\mathrm{nuc}}) A_{1l}^2 \cos\left(\frac{tA_{0l}}{2}\right)\cos\left[2\pi\nu_{\mathrm{nuc}}(t+NT)\right], \tag{S28}$$

Specializing Eq. S28 to the center of a lobe with weight $\Gamma$, we have $\nu_{\mathrm{nuc}}NT \in \mathbb{Z}$ and $\zeta_N(T\nu_{\mathrm{nuc}}) = \Gamma$, which results in Eq. 16 of the main text. Furthermore, direct evaluation of Eq. S27 at $t=0$ for a single $I=1/2$ nucleus gives $C(0) = \sin^2(NT\Gamma A_1/2)$, which is the form used in Section VI of the main text.

## VII. NUCLEAR SPIN SENSING SIMULATIONS

### A. $^{13}$C Sensing

To simulate the $^{13}$C sensing measurements [Fig. 4(b,d,h) of the main text], we model the electron spin coupled only to $^{13}$C nuclei on the OX063 molecule. These differ from the $^1$H sensing measurements in two respects: 1) only a small fraction of the 52 carbon sites are occupied at natural abundance, and 2) several carbon sites have sufficiently strong hyperfine couplings ($A_{\mathrm{iso}}, A_{\mathrm{aniso}} \gtrsim 10$ MHz) that the effective-Hamiltonian treatment of Section V is not accurate. We therefore simulate the full time-dependent Hamiltonian under the experimentally applied control waveform and

average the resulting signal over random realizations of the molecular orientation, the $^{13}$C isotope distribution, as well as the electron Rabi frequency distribution and inhomogeneous lineshape.

We label the $N_C = 52$ carbon sites on the molecule by $l \in \{1, \ldots, N_C\}$. A $^{13}$C isotope configuration is represented by a binary vector $\boldsymbol{\eta} \in \{0, 1\}^{N_C}$, where $\eta_l = 1$ indicates that site $l$ is occupied by a $^{13}$C isotope. At 1% natural abundance, we model the $\eta_l$ as independent Bernoulli random variables with probability $\mathbb{P}[\eta_l = 1] = 0.01$. For a given $\boldsymbol{\eta}$, we denote the occupied sites by $\mathcal{S}(\boldsymbol{\eta}) \equiv \{l \mid \eta_l = 1\}$.

For each occupied site $l \in \mathcal{S}(\boldsymbol{\eta})$, we calculate the hyperfine couplings using Eq. 13 and 14 of the main text:

$$A_{0l}(\hat{\mathbf{n}}) = A_{\text{iso},l} + A_{\text{aniso},l} \left[1 - 3(\hat{\mathbf{r}}_l \cdot \hat{\mathbf{n}})^2\right], \tag{S29}$$

$$A_{1l}(\hat{\mathbf{n}}) = 3 A_{\text{aniso},l} \ (\hat{\mathbf{r}}_l \cdot \hat{\mathbf{n}}) \sqrt{1 - (\hat{\mathbf{r}}_l \cdot \hat{\mathbf{n}})^2}, \tag{S30}$$

where $\mathbf{r}_l$ is the nucleus-electron separation vector for site $l$, $r_l \equiv |\mathbf{r}_l|$, and $\hat{\mathbf{r}}_l \equiv \mathbf{r}_l/r_l$. The values $A_{\text{iso},l}$ and $A_{\text{aniso},l}$ depend only on the site $l$ on the molecule. These coupling constants have been previously reported in the literature for the closely-related Finland trityl radical [6, 7], which differs from OX063 by the outermost substituents [8]. See Table 2 of Ref. [6] for a summary of the coupling constants. For the simulations here, we utilize these values for the carbon sites common to both the Finland and OX063 radicals, while using the point-dipole approximation $A_{\text{iso},l} = 0$ and $A_{\text{aniso},l} = \mu_0 \hbar \gamma_C \gamma_e / (4\pi r_l^3)$ for the remaining outer carbon sites. Here, $\gamma_C/(2\pi) = 10.7$ MHz/T is the $^{13}$C gyromagnetic ratio.

Denoting the electron Rabi frequency and resonance offset by $u$ and $\Delta\nu$, respectively, for each realization $(u, \Delta\nu, \boldsymbol{\eta}, \hat{\mathbf{n}})$ we construct the Hamiltonian

$$H(u, \Delta\nu, \boldsymbol{\eta}, \hat{\mathbf{n}}, t) = -2\pi \Delta\nu S^z + 2\pi u \ \mathbf{a}(t) \cdot \mathbf{S} + \sum_{l \in \mathcal{S}(\boldsymbol{\eta})} \left( \left[-2\pi \nu_C + A_{0l}(\hat{\mathbf{n}}) S^z\right] I_l^z + A_{1l}(\hat{\mathbf{n}}) S^z I_l^x \right), \tag{S31}$$

where $\mathbf{a}(t)$ is the applied control waveform and $\tau_c$ is the total waveform duration. Here, we chose the transverse nuclear spin axes such that the transverse hyperfine field points along the $x$ axis for each nucleus. We then numerically compute the propagator

$$U_f(u, \Delta\nu, \boldsymbol{\eta}, \hat{\mathbf{n}}) = \text{Texp}\left[-i \int_0^{\tau_c} dt \ H(u, \Delta\nu, \boldsymbol{\eta}, \hat{\mathbf{n}}, t)\right]. \tag{S32}$$

For each realization we define the signal

$$C(u, \Delta\nu, \boldsymbol{\eta}, \hat{\mathbf{n}}) = \text{Tr}\left[S^z U_f^\dagger(u, \Delta\nu, \boldsymbol{\eta}, \hat{\mathbf{n}}) S^z U_f(u, \Delta\nu, \boldsymbol{\eta}, \hat{\mathbf{n}})\right] / \text{Tr}\left[(S^z)^2\right]. \tag{S33}$$

The simulation returns the average signal

$$\bar{C} = \frac{1}{N_{\text{real}}} \sum_{i=1}^{N_{\text{real}}} C(u^{(i)}, \Delta\nu^{(i)}, \boldsymbol{\eta}^{(i)}, \hat{\mathbf{n}}^{(i)}), \tag{S34}$$

where $u^{(i)}$, $\Delta\nu^{(i)}$, $\boldsymbol{\eta}^{(i)}$, and $\hat{\mathbf{n}}^{(i)}$ are drawn independently, with $u_i$ sampled from the measured Rabi distribution [Fig. 1(e) of the main text], $\Delta\nu^{(i)}$ sampled from a Gaussian lineshape corresponding to $T_2^* = 78$ ns, $\eta_l^{(i)}$ sampled from the Bernoulli distribution above, and $\hat{\mathbf{n}}^{(i)}$ sampled uniformly on the unit sphere.

Equation S34 was used to calculate the solid lines in Fig. 4(d,h) of the main text, with $N_{\text{real}} = 10{,}000$ realizations. For the XYXYd duration sweep [Fig. 4(b)], the simulated curve is stitched from two models: in the interval $T \in [1.6 \ \mu\text{s}, 1.8 \ \mu\text{s}]$ we use the full $^{13}$C simulations of Eq. S34, while outside this interval we use the effective-Hamiltonian $^1$H model described in the next section.

We now quantify the relative contributions of different nuclear sites to the $t = 0$ correlation signal. In the small-phase regime, Eq. (16) of the main text implies an approximately additive contribution $\propto A_{1l}^2$ from each site. However, since the effective-Hamiltonian description breaks down for $^{13}$C nuclei, the correlation signal is not guaranteed to admit a unique decomposition into independent single-site contributions. Nevertheless, we define an operational measure of 'site contribution' by selectively removing nuclei from the Hamiltonian in the above waveform simulations [Eq. (S34)]. To this end, we introduce a radius parameter $r$ and define a truncated isotope-configuration vector $\tilde{\eta}_l(r) \equiv \eta_l \ \Theta(r - r_l)$, where $\Theta$ is the Heaviside step function, and $r_l$ is the distance between the electron and the $l^{\text{th}}$ nuclear site. In other words, we retain only occupied sites within distance $r$ from the electron, and set all other sites to $^{12}$C (no nuclear spin). We evaluate the corresponding averaged signal

$$\bar{C}(r) \equiv \frac{1}{N_{\text{real}}} \sum_{i=1}^{N_{\text{real}}} C\left(u^{(i)}, \Delta\nu^{(i)}, \tilde{\boldsymbol{\eta}}^{(i)}(r), \hat{\mathbf{n}}^{(i)}\right), \tag{S35}$$

9

where, for each value of $r$, the realizations $\{u^{(i)}, \Delta\nu^{(i)}, \hat{\mathbf{n}}^{(i)}, \boldsymbol{\eta}^{(i)}\}_{i=1}^{N_{\text{real}}}$ are drawn independently from the same distributions as in Eq. (S34) and $\tilde{\boldsymbol{\eta}}^{(i)}(r)$ is obtained from $\boldsymbol{\eta}^{(i)}$ via the truncation above. The function $\bar{C}(r)$ is therefore a cumulative measure of how the $t = 0$ correlation builds up as progressively more distant nuclear sites are included in the simulation. We evaluate $\bar{C}(r)$ at $r = r_l^+$, just beyond the $l^{\text{th}}$ carbon site. Fig. S5(a) depicts the calculated $\bar{C}(r)$ as a function of $r$. The fractional contribution of the $l^{\text{th}}$ site is defined as $w_l \propto \left[\bar{C}(r_l^+) - \bar{C}(r_l^-)\right]/M_l$, with normalization $\sum_l w_l = 1$, where $M_l$ is the number of carbon sites at the same distance $r_l$ from the electron. The inset of Fig. 4(h) in the main text shows a color map of the calculated $w_l$.

### B. $^1$H Sensing

The $^1$H simulations presented in the main text are based on the analytic signal expressions derived in Section VI B, evaluated for the 63 protons on the OX063 molecule. We compute the direct XYXYd signal from Eq. S21 and the correlation-spectroscopy signal from Eq. S27. In both cases, $I = 1/2$ and the required hyperfine couplings $A_{0l}(\hat{\mathbf{n}}), A_{1l}(\hat{\mathbf{n}})$ are calculated using the point-dipole approximation. Similar to the $^{13}$C case, we average the analytic signal over 1,000 random realizations of the molecular orientation, the electron Rabi frequency distribution, and inhomogeneous lineshape. Note that the latter two enter through the filter function weights $\Gamma_k$. This procedure was used to calculate the solid lines in Fig. 4(c,f) of the main text. For the XYXYd duration sweep [Fig. 4(b)], this model is used outside the interval $T \in [1.6\ \mu\text{s}, 1.8\ \mu\text{s}]$.

To compute the fractional contribution of each $^1$H site to the $t = 0$ correlation signal shown in the inset of Fig. 4(g), we use the small-phase expression Eq. (16), for which the site dependence enters through the factor $A_{1l}(\hat{\mathbf{n}})^2 \propto (\hat{\mathbf{r}}_l \cdot \hat{\mathbf{n}})^2 [1 - (\hat{\mathbf{r}}_l \cdot \hat{\mathbf{n}})^2]/r_l^6$ [Eq. (S30)]. Averaging over molecular orientations yields the fractional site contributions $w_l \propto 1/r_l^6$, normalized as $\sum_l w_l = 1$. Figure S5(b) depicts the cumulative distribution $\bar{C}(r)/C_{\max}(r) = \sum_{r_l \leq r} w_l$ of the fractional signal contribution.

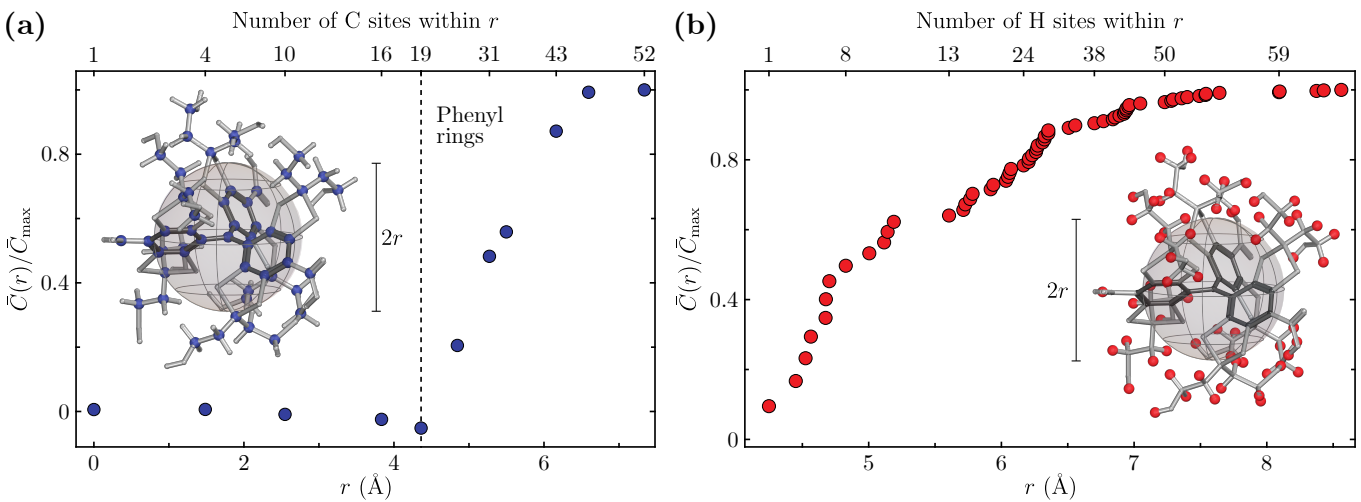

FIG. S5. Cumulative build-up of the $t = 0$ correlation signal versus cutoff radius for nuclei on OX063. (a) $^{13}$C: normalized cumulative signal $\bar{C}(r)/\bar{C}_{\max}$ obtained from the full waveform simulations by retaining only $^{13}$C sites within a sphere of radius $r$ around the electron. The upper axis shows the number of carbon sites within radius $r$. The dashed line marks the central carbon and phenyl-ring sites, which contribute negligibly due to strong hyperfine couplings. (b) $^1$H: corresponding normalized cumulative signal for the 63 protons, with the upper axis showing the number of proton sites within $r$. Insets: OX063 structure with the inclusion sphere (radius $r$) illustrating the truncation procedure. For each species, the 90% contribution threshold quoted in the main text is obtained by identifying the cutoff radius for which $\bar{C}(r)/\bar{C}_{\max} = 0.9$, and reading off the corresponding number of sites from the upper axis. For $^{13}$C, the quoted number excludes the 19 strongly coupled central-carbon and phenyl-ring sites (left of the dashed line), which contribute negligibly to the signal.

## VIII. SENSITIVITY CALCULATIONS

In this section, we summarize the details of the sensitivity calculations of Section VI of the main text. The signal-to-noise ratio (SNR) of statistical correlation measurements using the MAGGIC protocol has previously been analyzed

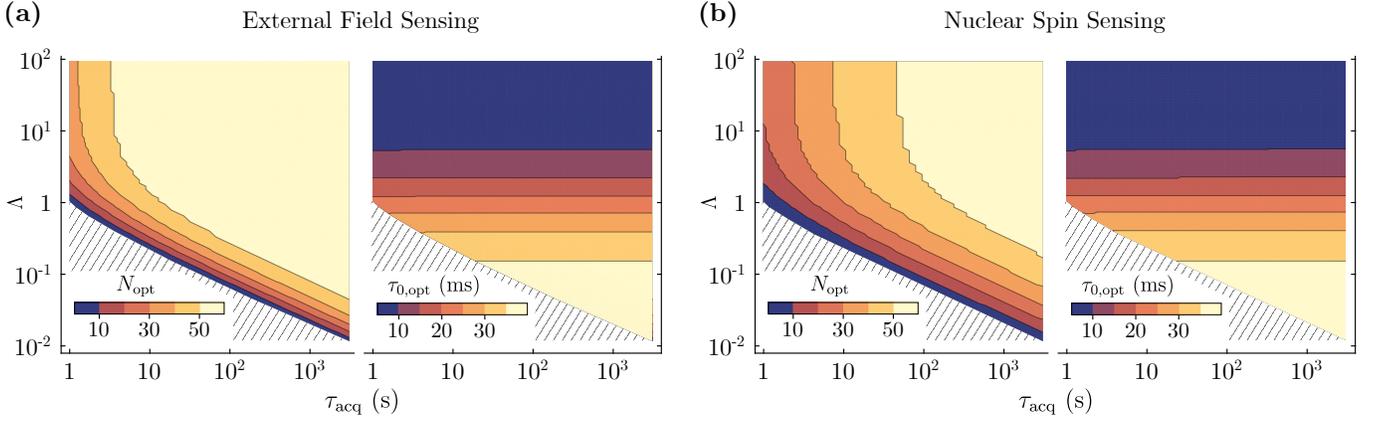

FIG. S6. Optimization parameters for the sensitivity calculations of Fig. 5 of the main text for external field sensing (a) and $^1$H spin sensing.

in detail in Section 4.2 of the Supplemental Material of Ref. [1]. Normalizing the expectation value of the bare electron signal to $\langle C \rangle = 1$, and assuming $\tau_c \ll T_1$, the variance of a single shot of the correlation measurement can be written as

$$\sigma^2\left(\Lambda, \frac{\tau_0}{\tau_m}\right) = \frac{4(\tau_0/\tau_m)^2}{\left(1 - e^{-\tau_0/\tau_m}\right)^4} \left[\alpha\left(\frac{\tau_0}{\tau_m}\right) + \beta\left(\frac{\tau_0}{\tau_m}\right)\frac{1}{\Lambda} + \frac{1}{4\Lambda^2}\right], \quad (S36)$$

where we have defined the auxiliary functions

$$\alpha(x) \equiv \left(1 - \frac{1 - e^{-x}}{x}\right)^2 \left[1 + \frac{(e^x - 1)^2}{4(e^{2x} - 1)[1 + e^x(x-1)]^2}\left((e^x - 1)^2\left(3 + e^{-2x}\right) + 4\left(1 - e^{-2x}\right)[1 + e^x(x-1)]\right)\right],$$

$$\beta(x) \equiv 1 - \frac{1 - e^{-x}}{x}. \quad (S37)$$

and $\Lambda \equiv \tau_m \mu^2 D^2 N_s \overline{G^2}/S_F$ is the readout efficiency.

The sensitivities in Fig. 5(a,b) are defined as unit-SNR detection thresholds for a fixed total acquisition time $\tau_{\text{acq}}$. For each pair $(\Lambda, \tau_{\text{acq}})$, the threshold is obtained by optimizing over the measurement-block duration $\tau_0$ and the number of XYXYd repetitions $N$. Throughout, we use the single-shot relative noise $\sigma(\Lambda, \tau_0/\tau_m)$ from Eq. S36, and we include the fixed per-shot overhead $\tau_{\text{ovh}} = 2$ ms arising from finite MAGGIC rise and fall times [9].

For a sinusoidal longitudinal field of amplitude $B_{\text{ext}}$ centered at a filter-function lobe with weight $\Gamma$, the single-shot SNR after averaging for time $\tau_{\text{acq}}$ is given by Eq. (17) of the main text. The minimum detectable field amplitude is then defined by the smallest positive solution to $\text{SNR}(B_{\text{ext}}, \tau_0, N) = 1$, i.e.

$$B_{\text{min}}(\Lambda, \tau_{\text{acq}}) = \min_{N, \tau_0} \; \frac{1}{NT\Gamma\gamma_e} \sin^{-1}\left[\sqrt{\frac{NT + \tau_0 + \tau_{\text{ovh}}}{\tau_{\text{acq}}}}\; \sigma\left(\Lambda, \frac{\tau_0}{\tau_m}\right) e^{NT/T_d}\right]. \quad (S38)$$

Similarly, using an XYXYd correlation spectroscopy sequence ($t = 0$), the SNR for detecting a single $^1$H spin via a pseudo-secular coupling strength $A_1$ is given by Eq. (18) of the main text, with the corresponding detection threshold being

$$A_{1,\text{min}}(\Lambda, \tau_{\text{acq}}) = \min_{N, \tau_0} \; \frac{2}{NT\Gamma} \sin^{-1}\left[\left(\sqrt{\frac{2NT + \tau_0 + \tau_{\text{ovh}}}{\tau_{\text{acq}}}}\; \sigma\left(\Lambda, \frac{\tau_0}{\tau_m}\right) e^{2NT/T_d}\right)^{1/2}\right]. \quad (S39)$$

For the optimizations in Eq. S38 and S39, we performed an explicit grid search over $\tau_0$ and $N$ and selected the smallest threshold value obtained on the grid. We evaluated the thresholds on a grid with $N \in \{1, 2, \ldots, 100\}$ and $\tau_0$ sampled using 344 logarithmically spaced points over $\tau_0 \in [2 \text{ ms}, 100 \text{ ms}]$. The lower bound on $\tau_0$ is set by the feedback-damped ringdown time of the oscillator. Grid points for which the argument of $\sin^{-1}$ exceeded unity were excluded, since they admit no unit-SNR solution [hatched regions in Fig. 5(a,b)]. We define the feasibility boundary as the boundary in $(\Lambda, \tau_{\text{acq}})$ beyond which no unit-SNR solution exists.

The optimized values of $N$ and $\tau_0$ are shown in Fig. S6 as a function of readout efficiency $\Lambda$ and acquisition time $\tau_{\text{acq}}$. The qualitative behavior of $N_{\text{opt}}$ depends on a trade-off between the amount of phase accrued and the electron spin decoherence under XYXYd (and to a lesser extent the added measurement overhead). This behavior can be understood by looking at two distinct regimes. Defining the $\sin^{-1}$ argument in Eq. S38 as $\Xi \equiv \sqrt{(NT + \tau_0 + \tau_{\text{ovh}})/\tau_{\text{acq}}}\,\sigma\left(\Lambda, \tau_0/\tau_m\right) e^{NT/T_d}$, for sufficiently large $\Lambda$ and $\tau_{\text{acq}}$, we have $\Xi \ll 1$ (linear response regime). This leads to

$$B_{\min} \simeq \frac{1}{\Gamma \gamma_e} \frac{\sigma\left(\Lambda, \tau_0/\tau_m\right)}{\sqrt{\tau_{\text{acq}}}} \frac{\sqrt{\tau_0 + \tau_{\text{ovh}}}}{NT} e^{NT/T_d}, \tag{S40}$$

where we also used the fact that $NT \ll \tau_0 + \tau_{\text{ovh}}$. In this case, Eq. S40 is minimized for $N_{\text{opt}} = T_d/T = 58$, which is independent of $\Lambda$ and $\tau_{\text{acq}}$. This sets the $N_{\text{opt}}$ plateau observed in Fig. S6 at large $\Lambda$ and $\tau_{\text{acq}}$.

The other edge case is near the feasibility boundary, where $\Xi \simeq 1$. In this limit, $\sin^{-1}(\Xi) \simeq \pi/2$ so $B_{\min} \simeq \pi/(2NT\Gamma\gamma_e)$, and minimizing $B_{\min}$ approximately reduces to maximizing $NT$ subject to the existence of a unit-SNR solution. Therefore, the approximate optimum is obtained by saturating the unit-SNR condition. Using $NT \ll \tau_0 + \tau_{\text{ovh}}$ gives $\sigma\left(\Lambda, \tau_0/\tau_m\right) e^{NT/T_d} \sqrt{\tau_0 + \tau_{\text{ovh}}/\tau_{\text{acq}}} \simeq 1$, i.e.

$$N_{\text{opt}} \simeq \frac{T_d}{T} \ln\left[\frac{\sqrt{\tau_{\text{acq}}}}{\sigma(\Lambda, \tau_0/\tau_m)\sqrt{\tau_0 + \tau_{\text{ovh}}}}\right], \tag{S41}$$

which decreases when $\Lambda$ or $\tau_{\text{acq}}$ are reduced, consistent with Fig. S6.

The optimized measurement duration $\tau_{0,\text{opt}}$ is primarily set by maximizing the MAGGIC readout SNR $\propto [(\tau_0 + \tau_{\text{ovh}})^{1/2} \sigma(\Lambda, \tau_0/\tau_m)]^{-1}$. Therefore, $\tau_{0,\text{opt}}$ is approximately independent of $\tau_{\text{acq}}$. For a small $\Lambda$, the measurement noise is dominated by thermal force noise, which incentivizes an increase in the measure block duration $\tau_{0,\text{opt}}$ for sufficient averaging in each measure block.

While the discussion above focused on external field sensing, the same arguments apply to the nuclear spin sensing case.

---


\AtBeginShipoutNext{\AtBeginShipoutDiscard}
\end{document}